\pdfoutput=1
\documentclass[a4paper,fleqn,usenatbib]{mnras}

 \usepackage{newtxtext,newtxmath}  

\usepackage[T1]{fontenc}
\usepackage{ae,aecompl}

\usepackage{graphicx,caption}			
\usepackage{amsmath}			
\usepackage{amssymb}		
\usepackage{mathtools}  		
\usepackage{natbib}
\setcitestyle{aysep{ }}
\usepackage{hhline}
\usepackage[normalem]{ulem}  
\usepackage{hyperref}			
\usepackage{booktabs}   		
\usepackage{stackengine}[2013-09-11]   
\usepackage{tabu}                 
\usepackage{longtable}        
\usepackage{dcolumn}        
\usepackage{subfig}
\usepackage[export]{adjustbox}
\usepackage{float}

\usepackage{syntonly}   
  
\newcommand{\HA}{H$\alpha$}	   				 
\newcommand{\MS}{\textnormal{M}$_{\odot}$}	    				 

\title[Radial distribution of SNe compared to SF tracers]{The radial distribution of supernovae compared to star formation tracers}

\author[F.M. Audcent-Ross et al.]{
\noindent Fiona M. Audcent-Ross,$^{1}$\thanks{E-mail: fiona.audcent-ross@icrar.org}
Gerhardt R. Meurer,$^{1}$
James R. Audcent,$^{2}$
Stuart D. Ryder, $^{3}$
\newauthor
O.I. Wong,$^{1,4}$
J. Phan,$^{2}$ 
A. Williamson,$^{5}$ and
J.H. Kim$^{6,7}$
\\
\textit{$^{1} $ International Centre for Radio Astronomy Research, University of Western Australia, 35 Stirling Highway, Crawley, WA 6009, Australia} \\
$^{2}$ University of Western Australia, 35 Stirling Highway, Crawley, WA 6009, Australia\\
$^{3}$ Department of Physics and Astronomy, Macquarie University, Sydney, NSW 2109, Australia\\
$^{4} $ ARC Centre of Excellence for All Sky Astrophysics in 3 Dimensions (ASTRO 3D), Australia\\
$^{5}$ Curtin University, Kent Street, Bentley, WA 6102, Australia\\
$^{6}$ Subaru Telescope, National Astronomical Observatory of Japan, 650 North A'ohoku Place, Hilo, HI 96720, USA\\
$^{7}$ Metaspace, 36 Nonhyeon-ro, Gangnam-gu, Seoul, 06312, Republic of Korea\\  
}
\date{Accepted xx. Received YYY; in original form ZZZ}

\pubyear{2019}

\begin{document}
\label{firstpage}
\pagerange{\pageref{firstpage}--\pageref{lastpage}}
\maketitle
\begin{abstract}

Given the limited availability of direct evidence (pre-explosion observations) for supernova (SN) progenitors, the location of supernovae (SNe) within their host galaxies can be used to set limits on one of their most fundamental characteristics, their initial progenitor mass.   We present our constraints on SN progenitors derived by comparing the  radial distributions of 80 SNe  in the SINGG and SUNGG surveys to the \textit{R}-band, \HA{}, and UV light distributions of the 55 host galaxies.  The strong correlation of Type Ia SNe with \textit{R}-band light is consistent with models containing only low mass progenitors, reflecting earlier findings.  When we limit the analysis of Type II SNe to apertures containing 90 per cent of the total flux, the radial distribution of these SNe  best traces far ultraviolet (FUV) emission, consistent with recent direct detections indicating Type II SNe have moderately massive red supergiant progenitors.   Stripped Envelope (SE) SNe have the strongest correlation with  \HA{} fluxes, indicative of very massive progenitors (M$_{*}  \gtrsim  ~20$ \MS{}).  This result contradicts a small, but growing, number of direct detections of SE SN  progenitors indicating they are moderately massive binary systems. Our result is consistent, however, with a recent population analysis suggesting binary SE SN progenitor masses are regularly underestimated.  SE SNe are centralised with respect to Type II SNe and there are no SE SNe recorded beyond half the maximum disc radius in the optical and one third the disc radius in the ultraviolet.  The absence of SE SNe beyond these distances is consistent with reduced massive star formation efficiencies  in the outskirts of the host galaxies.

\end{abstract}




\section{INTRODUCTION}
\label{sec:intro}

\bigskip
\bigskip
\bigskip
\bigskip
\bigskip

The number of supernovae (SNe) being detected has increased significantly in the past decade \citep[e.g. \citealt*{RN1479};][] {RN1410} due to the detection of fainter SNe  and expanding automated monitoring programs \citep[e.g.][]{RN1378,RN1725,RN1726,RN1868}. There is, however, only a small number of nearby SNe cases for which the SN progenitor has been directly identified \citep[see references in][]{RN1543,RN1563}.

Given limited direct evidence on the nature of SN progenitors, various researchers are using indirect methods to constrain key progenitor characteristics, such as mass and metallicity.   Examining both the host galaxy and the local environment in which SNe have occurred has, therefore, become important.   Stars, particularly massive ones, are believed to form in clusters \citep[e.g.][]{RN451,RN1848} and, given low typical velocity dispersions \citep*[e.g.][]{RN1370,RN1847}, massive SN  progenitors will therefore end their short lives in, or not far from, their birthplace environments.  Recent indirect methods used to study SN  progenitors include examining the radial distributions of the SN population \citep*[e.g.][]{RN1633, RN1562, RN1359}, analysing and modelling SN light curve behaviour \citep[e.g.][]{RN1747,RN1419} and dating the stellar populations near the SN location \citep[e.g.][]{RN1417,RN1473,RN1490,RN1498}.  This work uses the radial aperture analysis method (see Section \ref{sec:method}) to gain insights into the likely progenitor masses of key SN types.  

Supernovae fall into two broad categories:  Type Ia SNe and core collapse supernovae (CCSNe).  CCSNe occur at the end of the main sequence life of massive stars ($\textnormal{M}\gtrsim$ 8 $\textnormal{M}_{\odot}$) when exothermic fusion in the core ceases, leading to the rapid gravitational collapse of the core and subsequent explosive ejection of the outer layers of the star \citep{RN1372}. Either a neutron star or a black hole is formed, dependent on the stellar mass of the progenitor at the point of collapse \citep{RN1439}.      

SNe are classified according to their spectral properties \citep*[see][and references therein]{RN1813,RN1404,RN1840}, with Type II CCSNe having hydrogen  lines, unlike Types Ib and Ic.   The diversity of CCSNe types is believed to reflect  the extent of the progenitor's hydrogen envelope retained at the time of explosion.  Type Ib spectra contain helium lines, but Type Ic have neither hydrogen nor helium spectral features, having experienced the loss of their outer layers.  Immediately after exploding, a Type IIb SN exhibits spectral features similar to a Type II CCSN but  the hydrogen lines in the spectra disappear quickly (often within weeks) and the observed spectra is then typical of a Type Ib \citep*[e.g.][]{RN1625}.   Prompt observation of SN spectra is, therefore, essential for accurate distinction between these related SN types \citep[e.g.][]{RN1501}.  Type IIb SN progenitors are thought to retain as little as 0.01 \MS{} of their hydrogen envelopes at the time of exploding \citep[][]{RN1612}, explaining the short-lived hydrogen spectral features.  Stripped-envelope supernova (SE SNe: Types Ib, Ic, Ib/c  and IIb) are grouped together in this paper, as these spectrally-related objects are expected to have similar progenitor channels  \citep[e.g.][]{RN1642}.  

Type II CCSNe can also be classified according to the shape of their light curves in the weeks after going supernova; cases where the light curve drops linearly are classified IIL while the more common Type IIP feature a plateau phase.   There is, however, growing support for the view that Types IIL and IIP, as originally defined, are part of a larger continuous distribution \citep[see discussion in][]{RN1535}.  

For a star to terminate as a SE SN it must lose its outer layers of hydrogen (for a Type Ib) and, for Type Ic,  also its layers of helium.  See \citet*{RN1546} for a review of mass loss in SE SN progenitors.  This mass loss could arise from one of the following proposed mechanisms, or a combination of them: mass transfer to a nearby companion \citep[e.g.][]{RN1849,RN1771,RN1399}, ejection of a common envelope in  a binary system \citep*{RN1305},  precursor luminous blue variable (LBV) eruptions \citep*{RN1772,RN1470,RN1420,RN1762}, pulsation-driven superwinds in red supergiants \citep{RN1770}, or strong winds in Wolf-Rayet (WR) stars \citep{RN1764,RN1765}.

Beyond the Local Group, Type Ia SNe are the primary distance probes used by astronomers \citep{RN1679} and have played a pivotal role in the discovery of the accelerating expansion of the Universe \citep{RN1735}.  These extremely bright thermonuclear explosions are thought to arise from the destruction of  mass-accreting or merging carbon-oxygen white dwarfs \citep*[WDs:][]{RN1330, RN1752}.   Type Ia SNe have strong ionized silicon lines in their spectra and, unlike CCSNe,  have no hydrogen or helium lines, indicating that the progenitors are not main sequence stars.  

Despite their importance, the progenitors for Type Ia remain subject to conjecture, with no direct detections to date and several possible progenitor models under consideration \citep*[see][for a recent review]{RN1603}.  In the single degenerate (SD) model the WD progenitor gains mass though accretion from a non-degenerate companion (either a red giant or main sequence star) and, upon reaching the Chandrasekhar limit \citep*[M$_{\textnormal{Ch}} \approx 1.4$ \MS:][]{RN1312, RN1319}, explodes \citep*[e.g.][]{RN1310,RN1744,RN1746}.  Different variants on the SD model include a range of masses and evolutionary states for the companion and a range of WD progenitor masses: below, at, or above M$_{\textnormal{Ch}}$  \citep[e.g. \citealt*{RN1705,RN1704};][]{RN1734,RN1739,RN1661,RN1662}.  In the double-degenerate (DD) model,  two low mass WDs in a binary system merge due to gravitational interaction \citep*[][]{RN1314,RN1315}, taking the combined mass over the Chandrasekhar limit \citep*[][]{RN1743}.   

In order to constrain progenitor properties by location within the host, it is important to start with a well selected sample of potential host galaxies.  The potential host selection criteria should be homogeneous and well stated so that the inevitable biases can be identified.  \citet{RN1359}, for example, note that their heterogeneous sample has an increased bias towards brighter SNe and brighter galaxies. 

Surveys based on optically selected potential hosts are estimated to miss up to 20 per cent or more of SNe  in the local Universe \citep[e.g.][]{RN1497, RN1861}, with increasing fractions missed at higher redshifts due to high extinction from dust.   Extinction can be extreme, for example, in the luminous infrared galaxies (LIRGs) that dominate star formation at higher redshifts   \citep{RN1858}.  Locally, dust in the bars and bulges of galaxies will preferentially obscure optically dim SNe \citep[\citealt*{RN1387}; \citealt{RN1382};][]{RN1447}.  Resolving new point sources arising from SN events in or near the very luminous centres of galaxies can also be difficult, especially where there is a strong flux gradient \citep[e.g.][]{RN1388}.   Saturation of detectors by nuclei can be problematic and can cause central SNe to be missed at optical wavelengths \citep[e.g.][]{RN1858}. 
  
Galaxy-targeted surveys are biased against low mass and low luminosity galaxies \citep{RN1382}, despite them making  an important contribution to the total star formation occurring in the local Universe \citep{RN1714}. 

\captionsetup[subfigure]{labelformat=empty}
\begin{figure*}
\subfloat[][]{\includegraphics[width=0.41\linewidth,valign=t]{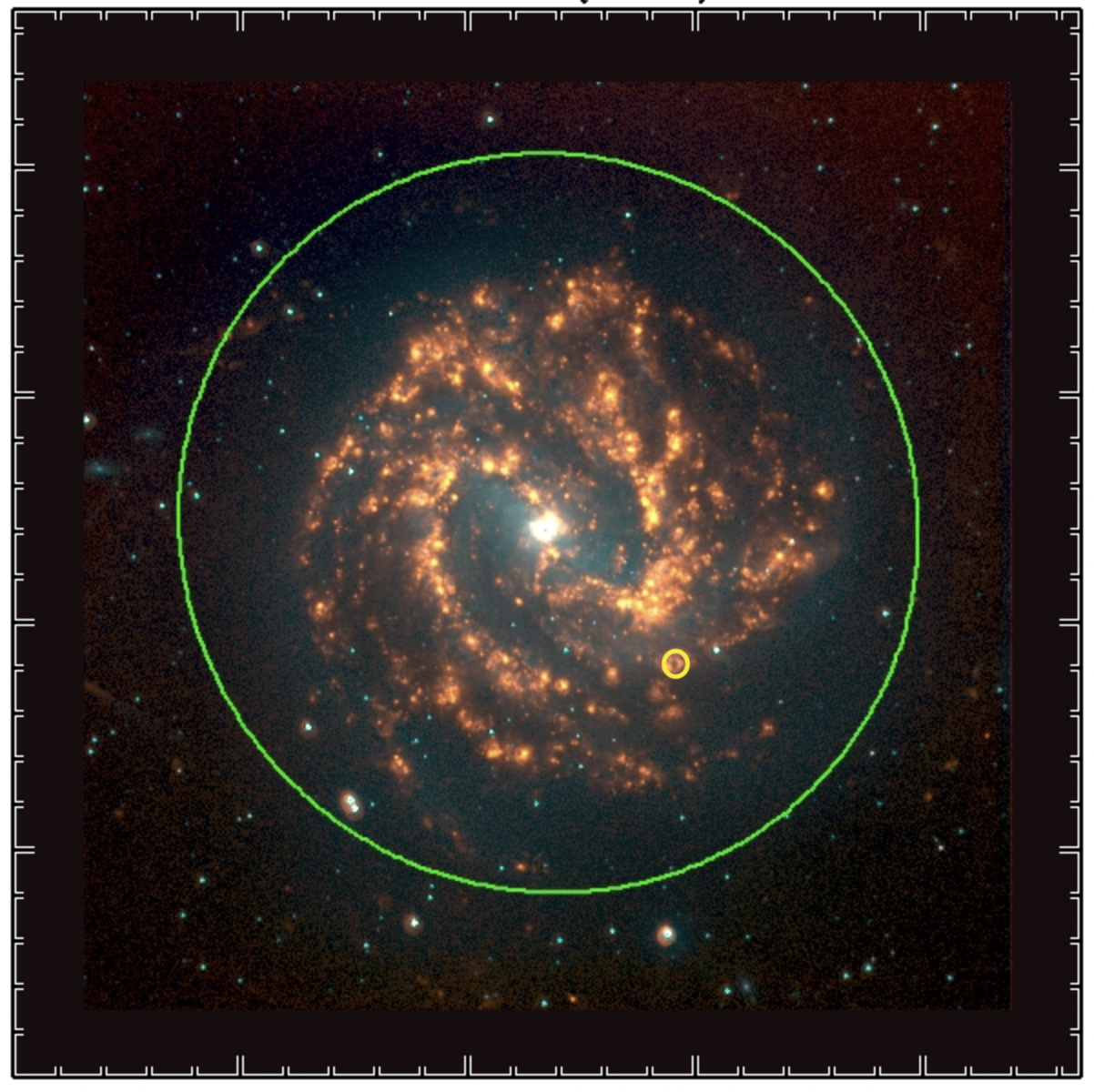} \label{fig:j1337_29}}
 \subfloat[][]{\includegraphics[width=0.48\linewidth,valign=t]{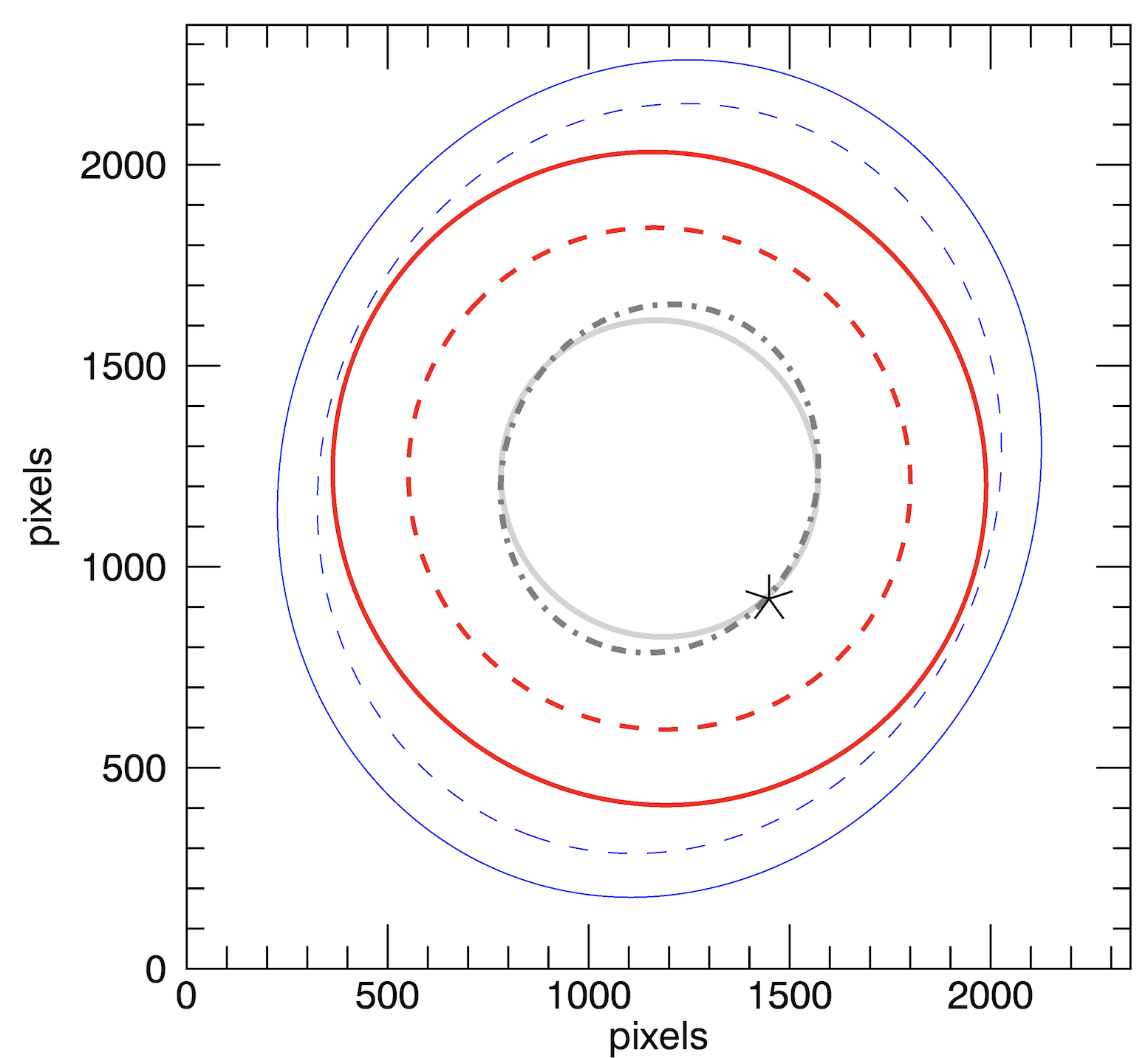} \label{fig:1983n_ap}}
 \caption{SN 1983N and host galaxy  J1337-29 (M83).  \textit{Left:} A three-colour image of J1337-29 with a yellow open circle indicating the location of SN 1983N inside the elliptical \textit{R}-band flux aperture \citep[of radius $\textnormal{R}= \textnormal{R}_{\rm max}$,][]{RN1779}.   \textit{R}-band flux is displayed in blue, narrow-band \HA{} (before continuum subtraction) in green  and net \HA{} (narrow-band \HA{} post continuum subtraction) in red.  North is up, east is to the left and the minor tick marks are 100 pixels (43") apart.   \textit{Right}: SN 1983N is indicated with a star and the \textit{R}-band and FUV $\textnormal{R}_{\rm max}$ apertures for the host galaxy are indicated with thick red and blue ellipses, respectively.  The concentric apertures containing 90 per cent of the \textit{R}-band and FUV fluxes are indicated with red and blue dashed ellipses, respectively.  The thick light grey solid line indicates the SN-enclosing aperture concentric with optical apertures and the thick dash-dot dark grey line marks the SN-enclosing aperture concentric with FUV apertures.  Note that the optimal SUNGG UV and SINGG optical apertures were determined independently, generating very similar, but not always identical, apertures (size, ellipticity and position angle).  Differences in ellipticity and position angle are generally small and immaterial, as seen here.  For an overview of the differences in aperture sizes see Figure \ref{fig:histo_r90opt_r90uv}. } \label{fig:1983n_2panels} 
\end{figure*}

Here we examine the constraints we can place on SN progenitors using a sample of potential host galaxies selected by atomic hydrogen (\textsc{Hi}) content in the very local Universe, and observed with star formation tracers in the optical and ultraviolet.  Specifically, our base of potential hosts is the galaxies observed for the Survey for Ionised Neutral Gas Galaxies (SINGG;  \citealt{RN41}), and the Survey of Ultraviolet emission in Neutral Gas Galaxies (SUNGG, Wong et al. in prep).  These surveys are discussed further in Section \ref{subsection:singg}. 

The paper is organised as follows: Section \ref{sec:method and sample section} presents our method, starting with the rationale behind it in Section \ref{sec:method_rationale}. Section \ref{sec:method and sample section} then explains the radial aperture analysis method  in detail and outlines the two surveys and the SN sample used in this work.  Section \ref{sec:Results} shows that the radial distribution of SNe is consistent with Type Ia progenitors having low mass companions; Type II progenitors have masses at the lower end of the mass range traced by FUV; and SE SNe originate from high mass stars.  These results are discussed further in Section \ref{sec:Discussion}.  We present our conclusions in Section \ref{sec:Conclusions}.

A Hubble constant of $H_0= $70 ~km s$^{-1}$ Mpc$^{-1}$ and cosmological parameters for a $\Lambda$CDM cosmology of $\Omega_{0} = 0.3$ and $\Omega_{\Lambda} = 0.7$ have been used throughout this paper.

\section{The radial aperture analysis method and our sample}
\label{sec:method and sample section}

\subsection{The rationale}
\label{sec:method_rationale}

Within late type galaxies the distance from an object to the galactic centre is a useful proxy for the likely local environment of that object.  Stars located centrally are, for example, more likely to be associated with the central bulge, normally comprising older, higher metallicity stellar populations \citep[e.g.][]{RN1816,RN1820,RN1817}.  Conversely, stars located further from the galactic centre are more likely to be in the stellar disc and to be associated with the younger stellar populations dominating the galaxy's spiral arms \citep*[e.g.][]{RN1819}. Radial aperture analysis compares the radial locations of SNe to the radial fluxes of host galaxies, giving insights into which stellar populations host SN progenitors.  Radial flux distributions can be impacted by internal and external factors, however, and these need to be considered when interpreting results. The actions of a strong bar, for example, can reduce local massive star formation \citep[e.g.][]{RN1429} and minor mergers or interactions can trigger considerable localised star formation \citep[e.g.][]{RN1822,RN1823,RN1821}, significantly altering radial flux distributions.

\subsection{The method}
\label{sec:method}

Radial aperture analysis  \citep{RN1306, RN1359} uses the total detectable optical and UV fluxes for a given galaxy using a curve of growth approach. Galaxy fluxes are measured within concentric elliptical apertures, centred on the galactic centre, out to the maximum radius at which light is detected above sky levels \citep[for further detail  see][]{RN41}.  The elliptical aperture enclosing the total detectable flux of the host galaxy (having a radius R $=$ R$_{\textnormal{max}}$; see \citealt{RN1779}) is then scaled down to determine the smaller, concentric ellipse that intersects with the individual SN (see Fig. \ref{fig:1983n_2panels}).  The ratio of the flux contained within the SN-enclosing ellipse to the total flux of the galaxy provides a useful measure of the supernova's centralisation.  A SN located at the host galaxy's centre has a radial enclosed flux fraction of zero, while a SN located at R$_{\textnormal{max}}$ has a radial enclosed flux fraction of 1.  Here we use four different radial flux distributions to derive a cumulative distribution function (CDF) of the flux contained within SN locations. Specifically, the four fluxes used are the \textit{R}-band continuum  and \HA{} from SINGG and the near and far ultraviolet (hereafter NUV and FUV) from SUNGG. A 1:1 CDF is indicative of an excellent radial tracer for the SN type under consideration.

This work utilises CDFs derived over two ranges; firstly, over the entire radial profile of the galaxy (R$_{\textnormal{max}})$ and, secondly, out to R$_\textnormal{{90}}$, where only 90 per cent of the total flux is contained.  Using R$_\textnormal{{90}}$ allows the bulk of the light to be assessed, while ignoring the outer regions where star formation can be significantly different due to local environmental factors.  Large changes in local star formation efficiency (SFE, the ratio of star formation rate relative to the available gas supply) in the outer disc may reflect areas where local gas surface densities fall below critical star formation thresholds \citep{RN433,RN1780}, or changes in disc stability \citep*[e.g.][]{RN1777}, accretion of cold gas from outside the disc \citep[][]{RN1808,RN1782}, or possible changes in the IMF (see Section \ref{subsect:centralisation of SE SNe}).   Sufficiently large changes in SFE will generate a break in the typically exponential form of a spiral galaxy's surface brightness profile \citep{RN1793,RN1787,RN1786}.  In SINGG, R$_\textnormal{{90}}$ typically lies just beyond where breaks in the basic exponential form occur \citep[][break radius R$_{\textnormal{B}}$ $\sim$ 0.75 R$_{90}$]{RN1786,RN1779}.  

\textit{R}-band, \HA{}, NUV and FUV CDFs are used as these fluxes are sensitive to different but overlapping mass ranges of stars, allowing us to constrain possible mass ranges for SN progenitors.  \HA{} is an excellent direct tracer of very high mass  star formation (M$_{*}  \gtrsim  ~20$ \MS{}), as only the most massive, short-lived ($t < 10$ Myr) O-type stars, are able to give rise to the photoionization of H\textsc{ii} regions, leading to recombination \HA{} emission.   \textit{R}-band is a good tracer of stars, covering the entire mass spectrum down to $\leq 1 \textnormal{M}_{\odot}$.  UV emission arises from both O- and B-type stars.  It is a useful indicator, therefore, of recent star formation of high mass stars (M$_{*} \gtrsim 3$ \MS{}).  For further explanation on the mass sensitivities of \HA{} and FUV emission see Section 3.1 of \citet{RN322}.

\begin{figure*}
\hspace*{-1.0cm}
\centering
  \includegraphics[width=0.85\linewidth,valign=t]{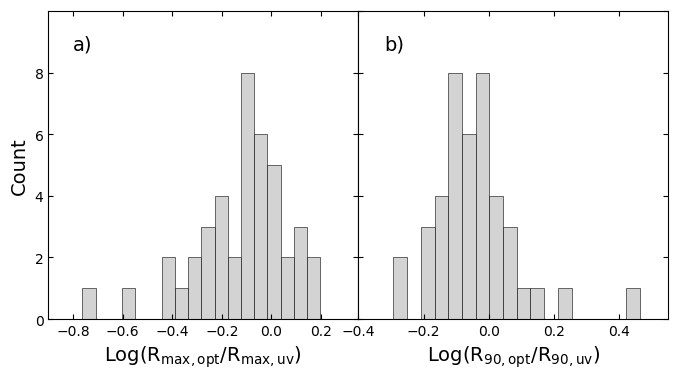} 
  \vspace*{0.5cm}
    \caption{Comparison of SINGG (optical) and SUNGG (UV) elliptical  flux measurement apertures at: (a) R$_{\textnormal{max}}$  and (b)  R$_{\textnormal{90}}$.  The 42 SNe host galaxies common to SINGG and SUNGG include many with an extended UV (XUV) disc  in comparison to their optical disc   \citep[][]{RN420}.  }   
\label{fig:histo_r90opt_r90uv}

\end{figure*}

\begin{figure}
\centering
\hspace*{-1.0cm}
 \includegraphics[width=0.85\linewidth,valign=t]{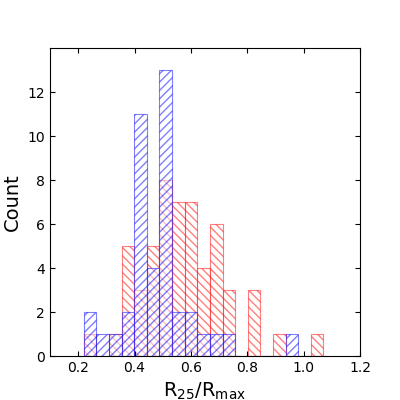}
   \vspace*{0.5cm} 
    \caption{Comparison of SINGG \textit{R}-band (red) and SUNGG FUV (blue) flux measurement apertures (R$_{\textnormal{max}}$) with R$_{25}$, the isophote at which the surface brightness of the host galaxy reduces to 25.0 mag per square arcsecond (measured in the \textit{B}-band).  The 55 SNe hosts have optical and, where applicable, UV apertures that are consistently larger than R$_{25}$,  with median R$_{25}$/R$_{\textnormal{max}}$ ratios of 0.55 and 0.49, respectively.  R$_{25}$ values have been obtained from NED; if multiple measurements were available the first-listed value in NED has been used. } 
\label{fig:r25flux}
\end{figure}

\subsection{An inclination cut}

Assuming circularity of disc galaxies, elliptical apertures arise from the inclination of the host galaxy with respect to the plane of the sky.  Line of sight issues are most extreme for galaxies with a large inclination value. Fluxes are measured using concentric elliptical apertures, regardless of the host's inclination. Light from highly inclined galaxies generally require very large corrections for extinction due to dust and, as a result, less luminous supernovae can remain undetected in  these galaxies  \citep{RN1497}. Only host galaxies meeting an axial ratio criterion (\textit{a/b}$ < 4$) were, therefore, used in our analysis.   

Line of sight considerations may bias apparent SN positions towards being more centralised than they really are.  In theory, a SN that appears to be centrally located could, for example, actually be part of the outer bulge stellar populations or the surrounding, more distant, halo.   This is unlikely to be a major issue, however, given very few SNe are located in galactic bulges \citep[e.g.][]{RN1718, RN1429,RN1428}.  In a review of 500 SNe in local non-disturbed galaxies \citet{RN1429}, for example, found all the CCSNe and the vast majority of Type Ia SNe were located in the disc, rather than in the bulge or halo components of the host galaxies. 

Despite the caveats outlined above, the radial distribution of historical SNe has been useful, identifying differences in  SN distributions by type and providing insights into both CCSNe and Type Ia progenitors \citep*[e.g.\citealt{RN1633};][]{ RN1627, RN1655, RN1657,RN1651,RN1359}.

\subsection{The SINGG and SUNGG surveys}
\label{subsection:singg}

SINGG and SUNGG are surveys of star formation in a sample of \textsc{Hi}-selected galaxies selected from the \textsc{Hi} Parkes All-Sky Survey \citep[HIPASS:][]{RN321, RN319, RN320}, using \HA{} and UV emission as tracers of star formation, respectively. SINGG \citep{RN41} observed 288 HIPASS sources revealing 466  galaxies  with \HA{} emission (at rest $\lambda = 6562.82$ \AA) using a variety of narrow band filters, and primarily a broad \textit{R}-band filter for continuum characterisation and subtraction.   Sources were selected to fully sample the  \textsc{Hi} mass function in order to obtain a complete view of star formation in the local Universe.  See \citet[][]{RN41,RN18,RN1714} and Meurer et al. (2020 in prep.) for further discussion.

SUNGG \citep[][]{RN381, RN257} uses GALEX       
NUV (central wavelength $\lambda_{c} =2273$ \AA) and FUV ($\lambda_{c} =1515$ \AA) observations of 418 unique galaxies with  \textsc{Hi} previously detected by HIPASS.  Only HIPASS sources meeting the axial ratio criterion \textit{a/b}$ <4 $ were used in SUNGG, as highly-inclined galaxies can experience severe extinction of their UV fluxes due to dust located in the plane of the host galaxy.  There are 231 HIPASS sources containing 320 star forming galaxies in common between the two surveys. 
   
The surveys have been used to measure the local star formation rate density, highlighting the importance of the contributions made from galaxy types that can be under-represented in optically-selected samples:  low  \textsc{Hi} mass, low luminosity and low surface brightness galaxies \citep{RN18,RN1714}.  Evidence of possible reduced massive star formation has been detected in low luminosity and low surface brightness galaxies as well as in the outskirts of galaxies \citep[][]{RN322, RN432,RN1309, RN1894}.

A common measure of a galaxy's radius is R$_{25}$, the isophote at which the surface brightness of the galaxy measured in the \textit{B}-band reduces to 25.0 mag per square arcsecond.  R$_{25}$ values for the host galaxies of our sample were extracted from the NASA/IPAC Extragalactic Database (NED),  using the first-listed value in NED, if multiple measurements were recorded.  SINGG and SUNGG apertures are, on average, double the  R$_{25}$ values  (see Fig. \ref{fig:r25flux}). The SINGG and SUNGG surveys are designed to measure the star formation rate density of the local universe and so the optical and UV elliptical apertures are set to ensure \textit{all} detectable fluxes from the target galaxy are included.  This was achieved using a curve of growth analysis to capture all the detectable optical and UV fluxes, respectively. As the optimal SUNGG UV and SINGG optical apertures are determined independently they are not always  identical in size, ellipticity and position angle (see Fig. \ref{fig:1983n_ap}).  The SINGG  apertures (both R$_{\textnormal{max}}$ and R$_{90}$) for the host galaxies are typically smaller than their SUNGG UV counterparts (see Fig. \ref{fig:histo_r90opt_r90uv}), consistent with many of these local galaxies having significantly extended UV discs  \citep[][]{RN485,RN420}.

\subsubsection{Non-stellar sources of \HA{} and FUV emission} 

Galactic \HA{} emission does not solely arise from massive star formation  \citep*[see][for a review of contaminating sources]{RN1719}; shocks \citep[e.g.][]{RN1814}, active galactic nuclei (AGN), planetary nebulae and supernova remnants (SNR), for example,  can generate \HA{} emission.  \citet{RN1719} measured the contribution of the observed SNR \HA{} emission to the total \HA{} flux for 25 very local galaxies that contain optically detected SNR, finding SNR generate 0.2--12.8 per cent of the total \HA{} fluxes.   Due to selection effects the measurements represent lower limits.  Two SINGG galaxies in our sample were examined by \citet{RN1719}; J2357-32 (NGC 7793) has a small measured SNR contribution of 1 per cent of total \HA{} emission from 27 SNR, while 296 known SNR in J1337-29 (M83) contribute 9 per cent of its total \HA{} emission.   The spatial distribution of SNR \citep[e.g.][]{RN1721,RN1724} in the 55 SINGG hosts could impact on the \HA{} CDF, but allowing for this complication is beyond the scope of this current work.  

Eight host galaxies (containing 13 SNe) are classified as Seyferts.  The AGN impact on the radial flux distributions of the hosts has been ignored in our analysis, even though AGN may generate considerable central \HA{} and FUV fluxes \citep[e.g.,][]{RN648,RN687}. However, the \HA{} flux emission from an AGN is typically dwarfed in comparison to emission at larger radii \citep [e.g.][]{RN648, RN687}.

\begin{figure}
\hspace*{-1.0cm}
\centering
 \includegraphics[width=0.85\linewidth,valign=t]{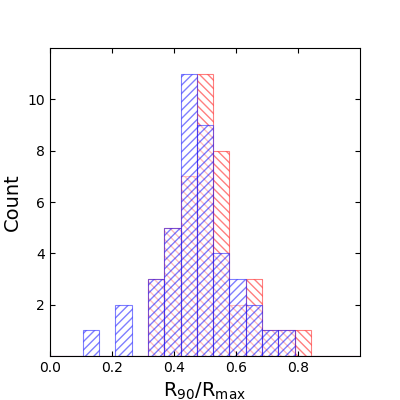}
    \vspace*{0.5cm} 
 
    \caption{Comparison of SINGG \textit{R}-band (red) and SUNGG FUV (blue) compactness (R$_{\textnormal{90}}$/R$_{\textnormal{max}}$) values  for the 42 host galaxies with both SINGG and SUNGG observations.  The low average compactness values reflect the significant centralisation of the \textit{R}-band and FUV fluxes (centred on R$_{\textnormal{90}}$/R$_{\textnormal{max}} \sim 0.5$). }   
\label{fig:compactness}
\end{figure}

\subsection{The supernova sample}
\label{subsect:Sample}

The IAU Central Bureau for Astronomical Telegrams (CBAT) listing\footnote{http://www.cbat.eps.harvard.edu/lists/Supernovae.html}, David  Bishop's "Latest Supernovae" website\footnote{http://www.rochesterastronomy.org/supernova.html}, and the Open Supernova Catalog \footnote{https://sne.space}\citep{RN1422}  were searched (all last accessed/downloaded 27th May 2019), by SN coordinates and host name, to identify recorded SNe that had potentially occurred within the galaxies of the SINGG and SUNGG surveys.  This approach was used to alleviate any potential bias against extreme outlying known SNe.  Targeted surveys using very limited fields of view will, however, generate a bias against the detection of extreme outlying SNe. See Section \ref{sec:intro} for more discussion on survey biases. 

All events in these catalogues known not to be a SN were ignored for this study.   The 466 star-forming  galaxies of SINGG contain 101 known SNe, including 18 of uncertain type.  The SNe of unknown type  are excluded from the radial aperture analysis.  Three host galaxies (J0953+01, J1445+01 and J1513-20) failed the axial ratio criterion (\textit{a/b}$ < 4$: see Section \ref{sec:method}) and the SNe located therein (1983E, 1983P, and 2002ds, respectively) are, therefore, also excluded from the analysis. 

The 80 SNe in our final sample represent all the most common types \citep[see][]{RN1665}, while rarer types, including Type IIn SNe \citep[e.g.][]{RN1388}, were not found.  There are 62 CCSNe and 18 Type Ia SNe and these are located at distances of 4--128 Mpc.  Type Ia SNe occur in both early and late types of galaxies, but the SINGG sample contains few early-type galaxies; the \textsc{Hi}-selection of the SINGG and SUNGG samples biases against early-type galaxies which typically have negligible or low levels of \textsc{Hi} and star formation.

\section{Results}
\label{sec:Results} 

\begin{figure*}
\centering
\hspace*{-0.7cm}
\includegraphics[scale=0.48]{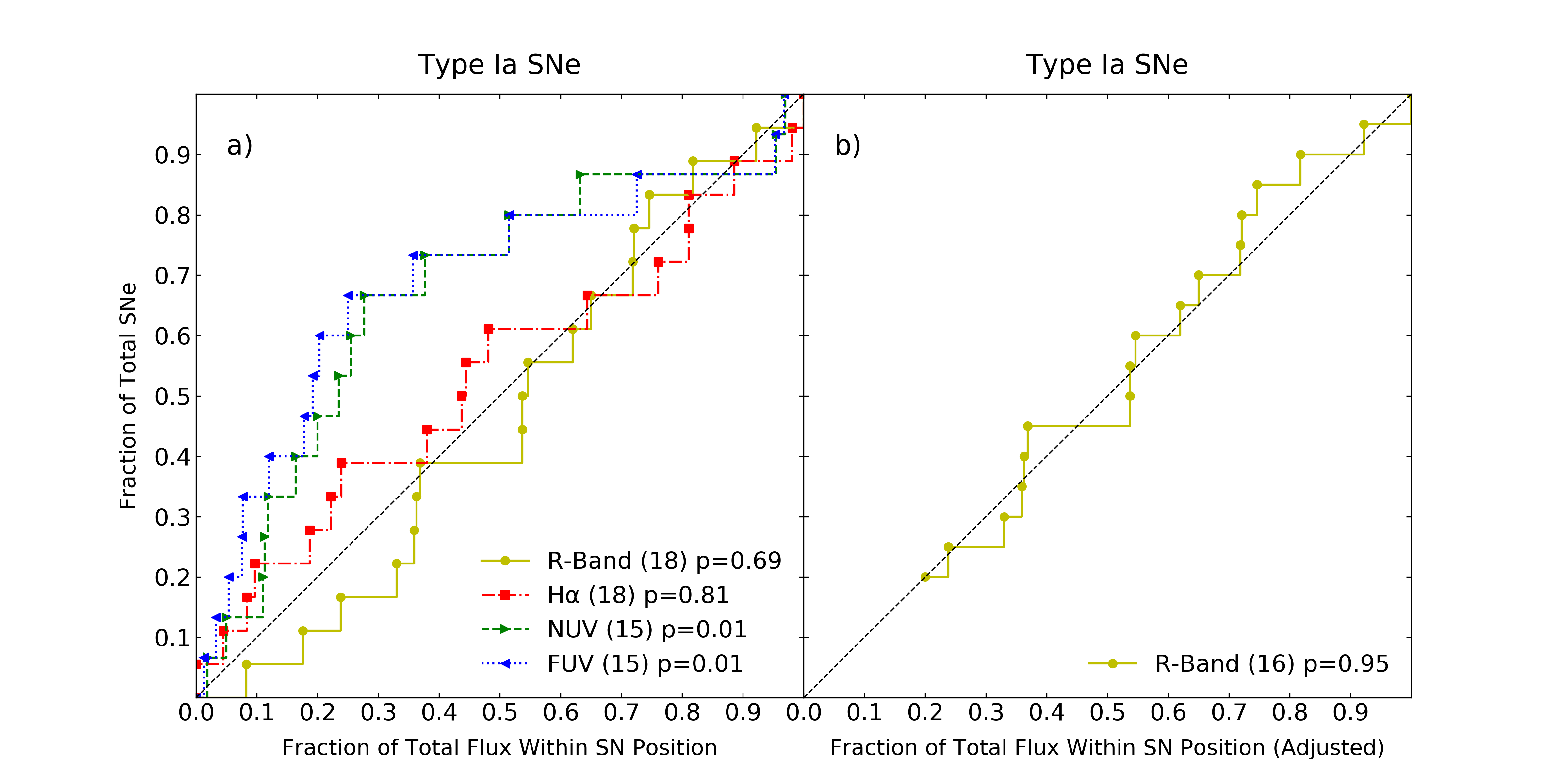}
 \caption{Cumulative distribution functions (CDFs) of the fluxes of the host galaxies interior to the Type Ia SNe positions (see Section \ref{sec:method} for method).  The \textit{R}-band, \HA{}, NUV and FUV CDFs are shown in yellow, red, green and blue, respectively, using the symbols shown in the key.   The number of SNe used in the construction of each CDF is also given in the key.  The diagonal dotted normal line is an ideal 1:1 CDF where the cumulative number of Type Ia SNe exactly traces the radial flux distribution.   The significance (p) of the one-tailed Kolmogorov-Smirnov (KS) testing is also listed, indicating the likelihood that the specified CDF is derived from the same population as the normal line. The Anderson-Darling test generates similar results; see Appendix \ref{appendix_histograms} for a summary. (a) CDFs show the radial distribution of all SINGG Type Ia SNe.  (b) Only Type Ia SNe occurring beyond the inner 20 per cent of the \textit{R}-band flux distribution are included in the  CDF here.  This CDF commences at (0.2, 0.2); the artificial start point reflecting the 1.00 slope of the best fit line (see Section \ref{subsect:central_deficit}) and the extent of observed central deficits \citep[e.g.][]{RN1737}.  As detailed in Section \ref{subsect:central_deficit}, bulges contribute significantly to central \textit{R}-band fluxes, while producing fewer Type Ia SNe per unit mass than discs.  The $\textnormal{p}=0.95$ result shows that the radial distribution of SINGG Type Ia SNe outside the central region is consistent with the \textit{R}-band light distribution. } 
\label{fig:ia_all_duo}
\end{figure*}

\begin{figure*}
\centering
\hspace*{-2.0cm}
\includegraphics[scale=0.36]{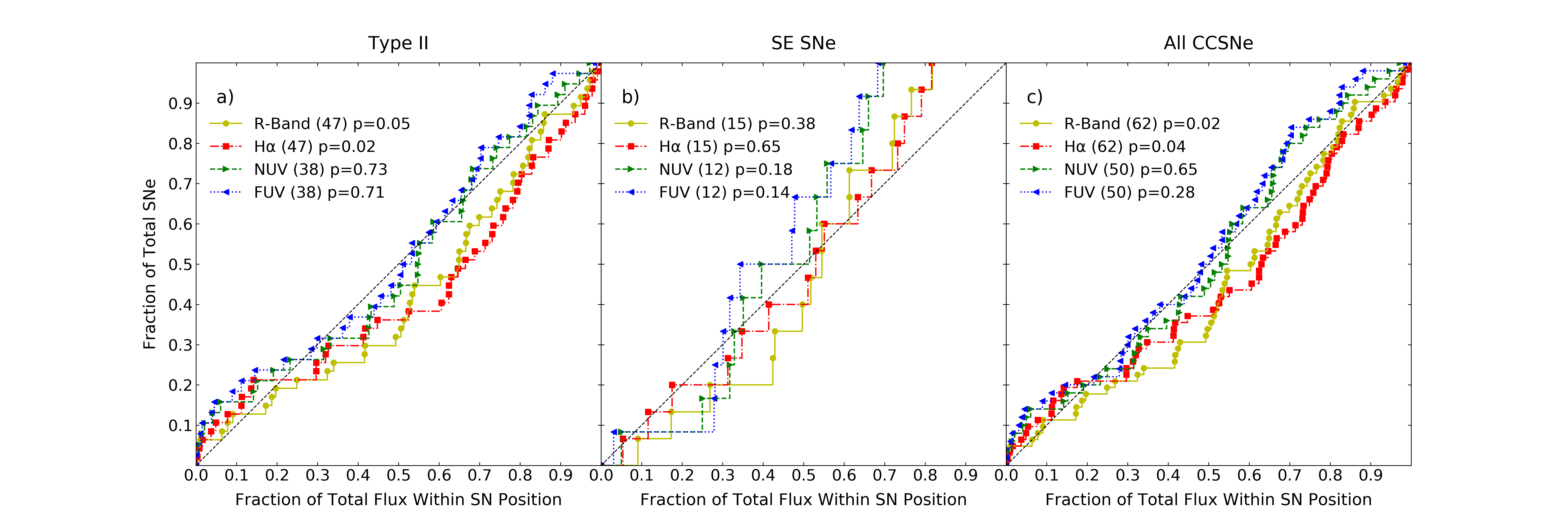}
\caption{CDFs for (a) Type II SNe, (b) stripped-envelope supernova (SE SNe: Types Ib, Ic, Ib/c  and IIb) and (c) the combined Type II and  SE SNe sample.   See Figure \ref{fig:ia_all_duo} for further description.}  

\label{fig:trioplot}
\end{figure*}

\subsection{Type Ia: A central deficit}
\label{subsect:central_deficit}

The radial distribution of Type Ia SNe in Figure \ref{fig:ia_all_duo}\textit{a} is best traced by the \textit{R}-band and \HA{} light distributions (Kolmogorov-Smirnov (KS) $\textnormal{p}=0.69$ and $0.81$, respectively).  The \textit{R}-band CDF  for Type Ia SNe generally lies below the normal line in the inner region (Fig. \ref{fig:ia_all_duo}\textit{a}), however, indicating a relative central deficit of SINGG Type Ia SNe with respect to \textit{R}-band flux (see also Fig. \ref{fig:fig9999}\textit{a}); only $\sim 11$ per cent of Type Ia SNe are located within the inner 20 per cent of  the total \textit{R}-band flux, for example.  This is consistent with observations by other researchers.  \citet{RN1737}, for example, used a much larger sample of 102 Type Ia SNe in star-forming galaxies and identified reduced numbers of Type Ia SNe within the inner 20 per cent of \textit{R}-band emission.  

In early work \citet*{RN1715} observed that SNe traced stellar luminosity, with a possible tendency to "avoid the nucleus"and the central deficit of Type Ia SNe in late type galaxies is well-known \citep*[e.g. \citealt*{RN1627};][\citealt{RN1737}]{RN1629, RN1657}.  In elliptical hosts the distribution of Type Ia SNe traces the overall galaxy light profile, albeit with a lower SNe rate per unit mass, and no central deficit is observed \citep[e.g.][]{RN1657, RN1651}.  Similarly, fewer Type Ia SNe per unit mass are produced in  bulges compared to the discs of spiral galaxies  \citep[e.g.][]{RN1737, RN1431}, resulting in the conclusion that the vast majority of Type Ia SNe are located in the disc, and not in the bulge or halo components, of their hosts \citep[e.g.][]{RN1718,RN1428}.  As the bulges of late type galaxies typically contribute approximately 30 per cent  of the total \textit{R}-band flux \citep[e.g.][]{RN1807}  but produce fewer Type Ia SNe per unit mass, the central  values of the \textit{R}-band CDF for Type Ia SNe are lower than the normal line, as observed in SINGG (Fig. \ref{fig:ia_all_duo}\textit{a}).

Excluding the two SINGG Type Ia SNe occurring inside the central, bulge-dominated, 20 per cent of  \textit{R}-band emission, Figure \ref{fig:ia_all_duo}\textit{b} shows that the radial distribution of the Type Ia population is strongly traced by the host galaxy's \textit{R}-band emission ($\textnormal{p}=0.95$).  Given the pivotal role of low mass WDs in all of the key models of Type Ia SNe generation \citep*[e.g.][]{RN1603}, the overall correlation of the Type Ia SN radial distribution with \textit{R}-band light distribution is expected.

Our results are consistent with the findings of \citet[][]{RN1737} who found the Type Ia SNe CDF followed \textit{R}-band fluxes. Their strongest correlation, however, was with \textit{B}-band fluxes; this wavelength is somewhat more sensitive to high mass stars than \textit{R}-band emission which traces a wide range of  stellar populations and stellar masses.  \citet*{RN1641} also found Type Ia SNe most closely followed a similar wavelength (g\textnormal{'}-band). The Type Ia SNe rate is highly dependent on global star formation, however,  and is, therefore, highly dependent on the host galaxy colour, with bluer galaxies having the highest Type Ia SNe rate per unit mass \citep*[e.g. \citealt{RN1325};][]{RN1324}. Type Ia SNe commonly occur in elliptical galaxies, including those with very little recent star formation \citep[e.g.][]{RN1326, RN1387, RN1325}, however, showing at least some progenitors are from older stellar populations. 

The radial distribution of Type Ia SNe  does not correlate with the UV flux distributions at a statistically significant level (see Fig. \ref{fig:ia_all_duo}\textit{a}: $\textnormal{p}=0.01$  for both NUV and FUV), in agreement with  \citet{RN1737}.  As UV is a useful tracer of moderate to high mass stars and, therefore, also a tracer of moderate emission time-scales \citep[e.g.][]{RN1881}, the divergence of the UV CDF from the dotted normal line in Figure \ref{fig:ia_all_duo}\textit{a} suggests that Type Ia SNe have progenitor systems from different mass and time-scale ranges.

\begin{figure*}
\centering
\hspace*{-0.6cm}
\includegraphics[width=1.05\linewidth]{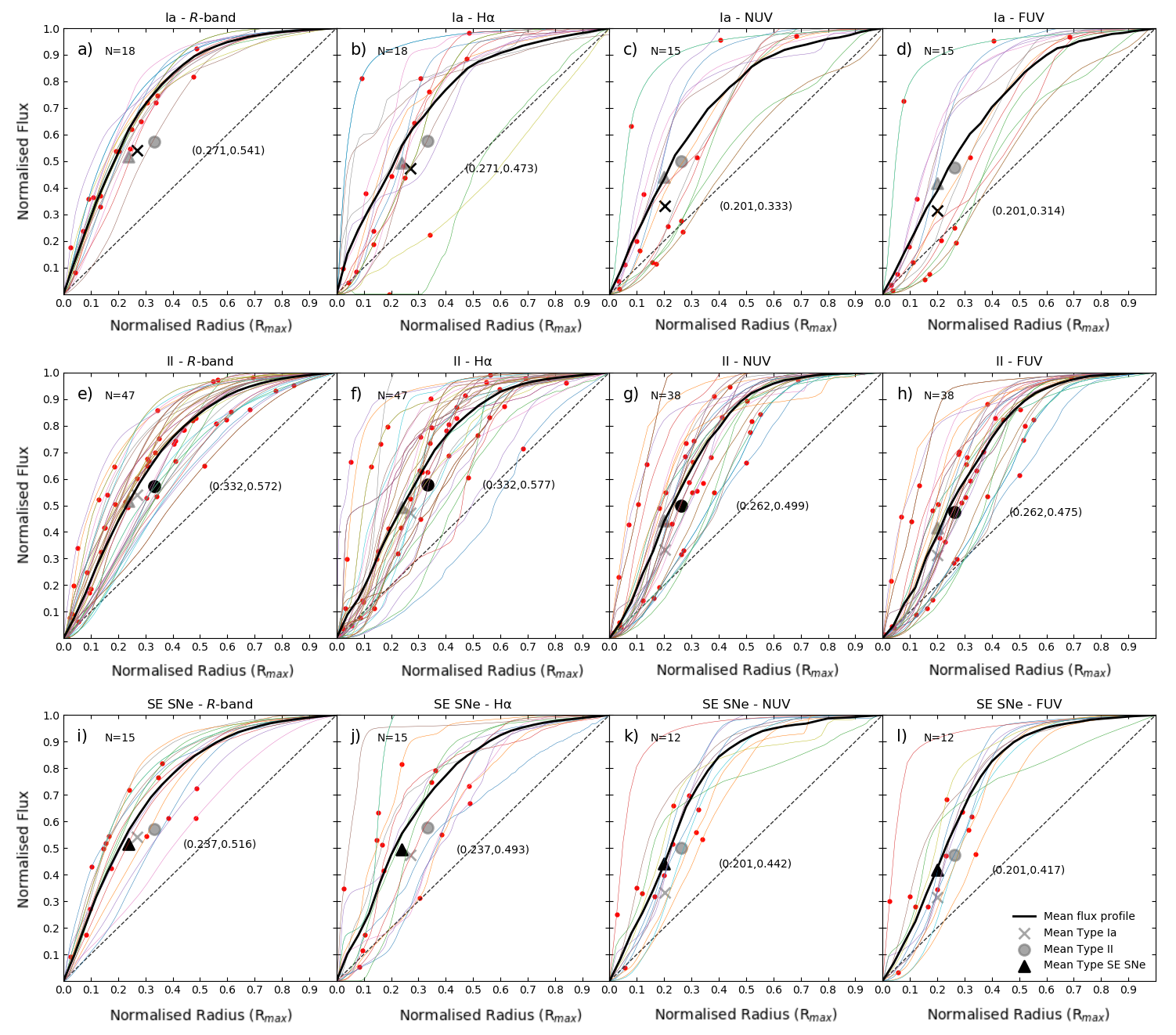}  
\vspace*{0.3cm}  
 \caption{Normalised radial flux profiles shown as separate differently coloured lines,  with SN locations indicated with small red circle(s):  Type Ia SNe (top row),  Type II SNe (middle row) and SE SNe (bottom row).  The average radial positions for Type Ia,  Type II and SE SNe  are marked on each plot using a black cross, circle and triangle, respectively, and these show the now well-known centralisation of SE SNe compared to Type II SNe.  To aid comparison the mean positions of the other SN types are shown with grey symbols.  The lack of SE SNe in the outer regions of the UV apertures is evident in panels \textit{k,l},  with no SE SNe observed in the outer $\sim30$ per cent of the UV light distribution, compared to the absence of SE SNe in the outer  $\sim18$ per cent of the \HA{} light distribution (panel \textit{j}).  See Sections \ref{subsubsect:sesn_centralised} and \ref{subsect:centralisation of SE SNe} for further discussion. }
 \label{fig:fluxcurves}
\end{figure*}

\bigskip

\subsection{Type II SNe}
\label{subsect:typeIIresults}

The radial distribution of Type II SNe  shows strong agreement with both UV fluxes ($\textnormal{p} = 0.73$ and $0.71$ for NUV and FUV, respectively)  and differs from the \textit{R}-band and \HA{}  light distributions at, or near, statistically significant levels: $\textnormal{p} =  0.05$ and $0.02$, respectively (see Fig. \ref{fig:trioplot}\textit{a} and Appendix \ref{appendix_histograms}). Type II SNe occur throughout the optical disc and examples can be found near or at R$_{\textnormal{max,opt}}$ (see also Fig. \ref{fig:fluxcurves}\textit{e,f}).  

The  \HA{} CDF for Type II SNe (see Fig. \ref{fig:trioplot}\textit{a}) has a central section with a reduced slope;  just over 10 per cent of Type II SNe are found in 30 per cent of the flux (between 30 to 60 per cent of the total flux - see also Fig. \ref{fig:fig9999}\textit{b}). Strong bars  are known to cause a notable suppression of massive star formation in early-type spirals \citep[see][and references therein]{RN1429} and this could contribute to this CDF feature.

No Type II SNe are found in the outer $\sim$ 30 per cent of the FUV apertures by radius (see Fig. \ref{fig:fluxcurves}\textit{g,h}).  This lack of Type II SNe  in the outskirts could arise from a combination of Type II SN progenitors having masses towards the higher end of the mass range traced by UV  and  weak star formation in outer discs, apparently with an Initial Mass Function deficient in the most massive stars \citep[e.g.][]{RN432,RN1309, RN1894}.  Metallicity gradients are unlikely to be important, with research showing that progenitor metallicity is not a significant factor in determining SN type \citep{RN1355}. 

The physical areas beyond R$_\textnormal{{90,UV}}$ and R$_\textnormal{{90,opt}}$ can be sizeable;  XUV discs are common \citep[][and see Fig. \ref{fig:histo_r90opt_r90uv}]{RN485,RN484,RN420} and \textit{R}-band fluxes and \HA{} emission are concentrated (see the normalised radial profiles in Figs \ref{fig:fluxcurves}\textit{e,f} and \ref{fig:compactness}).  As explained in Section \ref{sec:method}, performing  radial analysis within R$_{90}$ allows the bulk of the light to be assessed while ignoring the outskirts of the host galaxies, where local environmental factors may profoundly impact star formation.

\begin{figure*}
\centering
\hspace*{-1.8cm}
\includegraphics[scale=0.35]{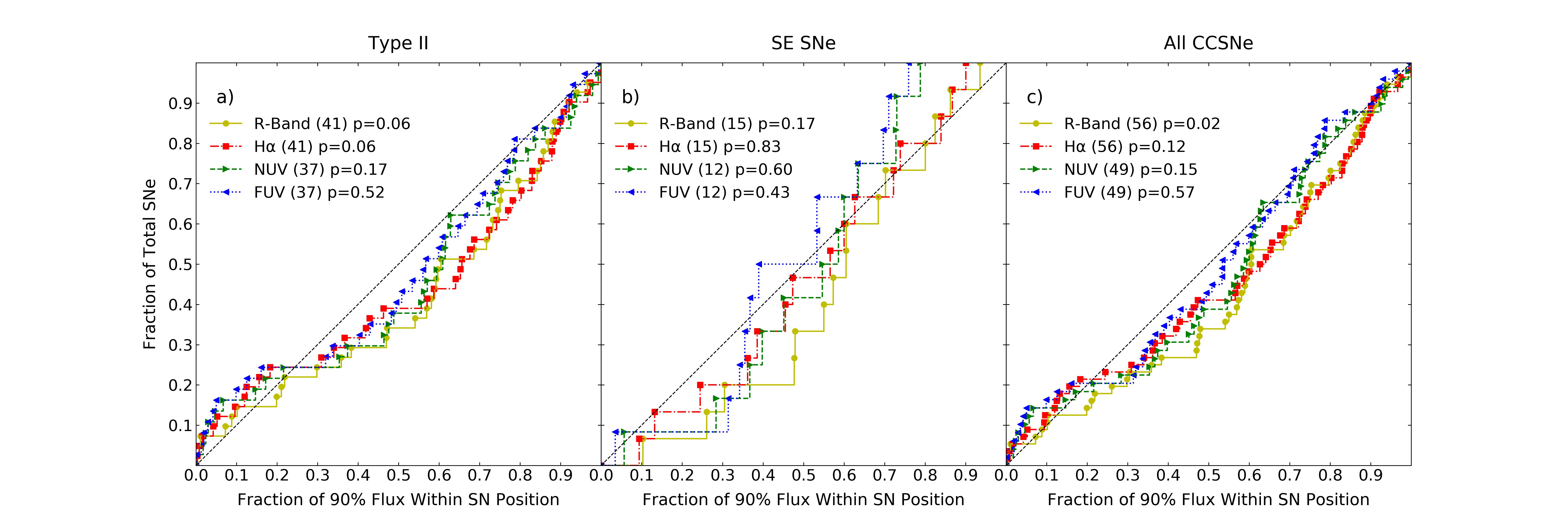}
\caption{CDFs for SNe located within  R$_{90}$: (a) Type II SNe, (b) stripped-envelope supernova (SE SNe: Types Ib, Ic, Ib/c  and IIb) and (c) the combined Type II and  SE SN sample.  See Figure \ref{fig:ia_all_duo} for further description.  }  
\label{fig:9007}
\end{figure*}

Type II SNe within R$_{90}$  are best traced by FUV ($\textnormal{p} = 0.52$: see Fig. \ref{fig:9007}\textit{a}) and deviate from \HA{}, although not at a statistically significant level ($\textnormal{p} = 0.06$).   Using a  much larger sample, \citet{RN1359} also observed the positive correlation of Type II SNe and FUV and the deviation from \HA{} emission.

The correlation of Type II SNe with FUV R$_{90}$ fluxes is consistent with growing evidence that most, if not all, of Type IIP SNe originate from single moderately massive (M$_\textnormal{{i}} \sim$ 8 - 16 \MS{}) red supergiants (RSG)  \citep[see][]{RN1545,RN1563}.  However, some research suggests Type II SNe cannot arise solely from single stars \citep[e.g.][]{RN1586}. There are only a few direct detections of Type IIL SN progenitors \citep*{RN1563} and these support slightly more massive progenitors for some Type IIL SNe   \citep[][but see \citealt{RN1561}]{RN1407,RN1421}.   

The low correlation  of the overall Type II SN distribution and \HA{}, tracer of very high mass stellar populations, is consistent with observational evidence and modelling that indicates that most very massive (M $\geq$ 18 M$_{ \odot}$) stars may collapse directly to a black hole \citep[to become "failed supernovae" e.g.][]{RN1775,RN1544,RN1487, RN1451,RN1543,RN1827,RN1353}.   \citet[][]{RN1353} confirmed the first case of a disappearing massive star using the Large Binocular Telescope to search for failed SN candidates \citep[N6946-BH1:][and see also \citealt*{RN1815}]{RN1467}.  The 25 \MS{} RSG progenitor experienced considerable mass loss of its outer layers before leaving signs of a newly formed black hole  \citep[e.g. late-time emission][]{RN1812}, without a CCSN explosion occurring.  The direct-to-black-hole evolution of massive stars could explain the  lack of high mass CCSN progenitors identified to date \citep[e.g.][]{RN1543, RN1353} and  the compact remnant mass function \citep*[see][]{RN1463}.  \citet{RN1390} found most (> 82 per cent) massive stars (M > 20 \MS{}) are in close binary systems and this can have a profound impact on the evolution of the progenitor \citep[see][and references therein]{RN1533}.  Over 70 per cent, for example, will exchange mass with a companion, and very close binaries may undergo a stellar merger before ultimately collapsing directly to form a black hole \citep{RN1533}.

\subsection{High mass progenitors for stripped-envelope SNe}

The radial distribution of SINGG SE SNe is best traced by the \HA{} light distribution ($\textnormal{p}=0.65$: see Fig. \ref{fig:trioplot}\textit{b}).  This result is consistent with SE SNe having high mass progenitors (M$_\textnormal{{i}}$ > 20 \MS{}) and with  earlier radial analysis research \citep{RN1306, RN1358,RN1357}.  The observed correlation of SE SNe with \HA{} fluxes increases when the radial analysis is restricted to  R$_\textnormal{{90}}$ ($\textnormal{p} = 0.83$: see Fig. \ref{fig:9007}\textit{b}), and UV fluxes improve as tracers ($\textnormal{p} = 0.60$ and $0.43$ for NUV and FUV, respectively).   It is well-known that  SE SNe are the CCSN subgroup most associated with high mass star-forming regions \citep[e.g][]{RN1620, RN1413}, and Ic SNe in particular \citep[e.g.][]{RN1358, RN1845}, implying a possible sequence of increasingly massive progenitors (II $ \rightarrow$ Ib $\rightarrow$ Ic). Some single massive (M$_\textnormal{{i}}$ > 25 \MS{}) WR star SE SN  progenitors have been successfully identified  or  modelled \citep[e.g.][]{RN1647,RN1396,RN1605,RN1595}, including the first Type Ic SN progenitor identification \citep[][]{RN1597,RN1806}.

There is growing evidence, however, that SE SNe may have at least two progenitor streams, spanning a wide range of masses \citep[e.g.][]{RN1305,RN1400,RN1568,RN1531,RN1612, RN1593}, but see \citet*{RN1498}.   Single massive star evolution models do not explain the observed ratio of SE SNe compared to Type II SNe, and a lower mass binary progenitor stream could explain the observed frequency of SE SNe \citep[e.g. \citealt*{RN1648}; \citealt{RN1547};][]{RN1755,RN1806}.  Recent modelling of light curves suggests that most SE SNe have low ejecta masses (M$_{\textnormal{ej}} = 1 - 5 $ \MS{}), whereas massive stars typically have high ejecta masses \citep[e.g.][]{RN1489, RN1612}.     The analysis of late-phase spectra of Type IIb SNe by \citet{RN1754} also indicates SE SNe can have moderately massive progenitors (typical M$_\textnormal{{i}}$ $ \approx 12 - 16 $ \MS{}), in line with many recent direct detections \citep[e.g. \citealt*{RN1606};][]{RN1501, RN1598}.       

UV fluxes  are not the best tracers of the radial distribution of the SE SNe within R$_\textnormal{{90}}$ (see Fig. \ref{fig:9007}\textit{b}).  If most SE SNe are produced by lower mass stars in binary systems, as postulated by \citet[][]{RN1396}, then a stronger correlation would be expected, but this is not seen.  \citet*{RN1498}, however,  suggests very massive stars (M$_\textnormal{{i}}$ > 30 \MS) produce the majority of SE SNe.   Using the \textit{Hubble Space Telescope} \citet*{RN1498} analysed the sites of 23 SE SNe, determining the age of the stellar populations remaining at the vicinity (< 150 pc) of the SN locations.  Deriving stellar population ages for the resolved stars local to the SN locations using color magnitude diagrams, they find SE SNe typically to be coeval with very young stellar populations.  This is indicative of much higher progenitor masses, either as single stars or in a massive binary system.   \citet{RN1498} finds higher extinction towards the SN sites than previously assumed, thus leading to higher progenitor mass estimates.  

\begin{figure*}
\centering
\hspace*{-1cm}
\includegraphics[scale=0.37]{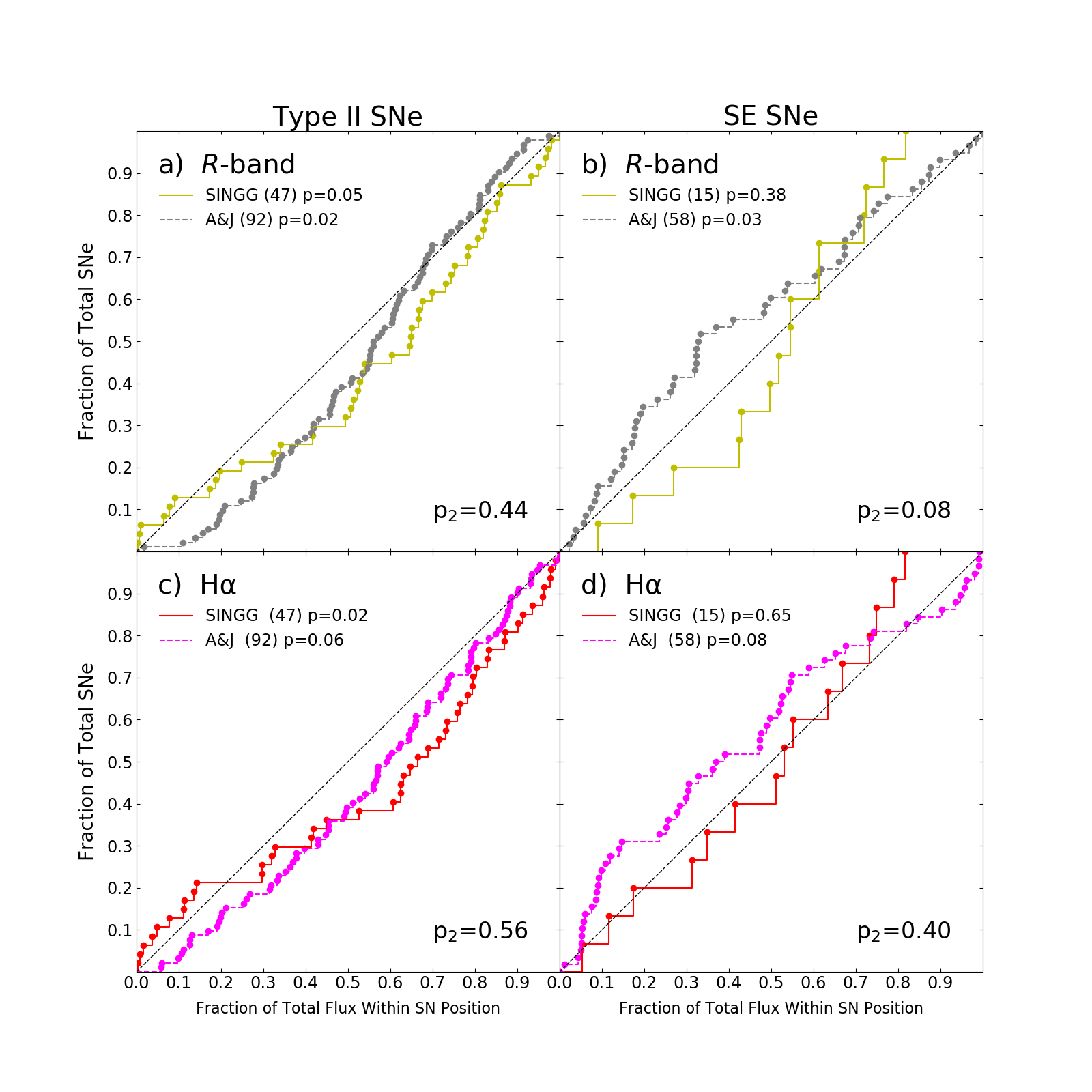}
\caption{Comparison of the \textit{R}-band and  \HA{} Type II SNe and SE SNe CDFs derived from the homogeneous SINGG dataset (solid lines in yellow and red for 47 Type II SNe and 15 SE SNe, respectively) with the \citet{RN1359} sample (dashed lines in grey and magenta for 92 Type II SNe and 58 SE SNe, respectively).  For consistency with SINGG, the \citet{RN1359} sample used here excludes ten SNe which are missing either \textit{R}-band or  \HA{} flux measurements.  As in earlier CDF figures, the diagonal dotted normal line reflects an ideal 1:1 CDF where the cumulative number of the SNe exactly traces the radial flux distribution.   The significance (p) of the one-tailed KS test is also listed, indicating the likelihood that the specified CDF is derived from the same population as the normal line. The significance of the two-tailed KS test (p$_{2}$) indicates the likelihood that the \citet{RN1359} and SINGG CDFs are derived from the same population.}  
\label{fig:ajvsingg}
\end{figure*}

We do not observe an excess of SE SNe near galactic centres (see Fig. \ref{fig:trioplot}\textit{b}), unlike some observers   \citep*[e.g.][]{RN1359,RN1426}, but perhaps this partially reflects the smaller size of our homogeneous sample.   The SE SNe in the  heterogeneous sample used by \citet{RN1359} are significantly more centralised than our sample, as can be seen in Figure \ref{fig:ajvsingg}.  Only two SE SNe (1997X and 2004dk) and two Type II SNe (1999em and 2004dg) occur in both samples; the SINGG \textit{R}-band and \HA{}  enclosed flux fractions for these SNe  are typically 6--12 per cent lower than the \citet{RN1359} values, consistent with our typically larger flux apertures set to capture all detectable fluxes (see Fig. \ref{fig:compactness} and Section \ref{subsect:Sample}). 

\subsubsection{SE SNe absent from galaxy outskirts}
\label{subsubsect:sesn_centralised}

The centralisation of SE SNe compared to Type II SNe noted in earlier work  \citep[e.g.][]{RN1562, RN1657, RN1359,RN1691} is evident (see Figs \ref{fig:trioplot}\textit{b}, \ref{fig:fluxcurves}\textit{i--l},\ref{fig:9007}\textit{b}).  There are no SINGG SE SNe observed in the radii containing the outer $\sim$ 30 per cent of the UV fluxes, nor in the outer $\sim$ 18 per cent of the optical fluxes (see Fig. \ref{fig:fluxcurves}\textit{i--l}).  In contrast, there are no Type II SNe in the outer five to ten per cent of the UV profiles (Fig. \ref{fig:fluxcurves}\textit{g,h}) while SINGG Type II SNe are located out to the full  \textit{R}-band apertures (Fig. \ref{fig:fluxcurves}\textit{e,f}).   \citet{RN1359} also observed that no Type Ic SNe occurred in the outer 20 per cent of their \HA{} and  \textit{R}-band flux distributions.  Possible explanations for these results are discussed in Section \ref{subsect:centralisation of SE SNe}.  Recall that the large SINGG and SUNGG apertures probe the full extent of detectable fluxes (Section \ref{subsection:singg}) and this work, therefore, examines the radial distribution of SNe well beyond the commonly used R$_{25}$ apertures \citep[e.g.][and see Fig. \ref{fig:r25flux}]{RN1892}.

\section{Discussion}
\label{sec:Discussion} 

\subsection{Low masses for Type Ia SNe binaries}
\label{subsect:typeia} 

\bigskip
\bigskip

The low likelihood that the SUNGG UV light  distribution and Type Ia SN radial distribution of the sample are related (Fig.  \ref{fig:ia_all_duo}\textit{a}) suggests that at least the majority of SINGG Type Ia SNe originate from binary systems containing only low mass stars. Galactic NUV emission is a tracer of star formation out to a few $\sim$100 Myr \citep[e.g.][]{RN322}, so the divergence is consistent with Type Ia SNe having older (> 100 Myr), low mass progenitors. 

Our results cannot distinguish between key progenitor models that involve only low mass stars. This includes the double degenerate model  \citep*[see review in][]{RN1495} and the single degenerate (SD) model  case with a low mass companion \citep*{RN1310}.  In the SD model  the progenitor, a main sequence star with a mass of M$_\star < $8 \MS, can evolve rapidly to form a WD of {$\sim $ 1.4 \MS}  and subsequently accrete sufficient material from a non-degenerate companion, taking the primary star to, or near, the Chandrasekhar limit, causing it to explode \citep{RN1330}. The SD model favours  an initial progenitor mass range of M$_\star \sim 1.8$--$3 $ \MS, while the moderately massive portion of the range (M$_\star \sim 3$--$8 $ \MS) is viewed as suboptimal \citep*{RN1310}.  \citet{RN1746} also suggest a potential range of companion masses (M$_\star \sim 0.7$--$6 $ \MS).   According to \citet[][]{RN1746} companions within the mass range can generate the high accretion rates required for a Type Ia SN  to occur and also explain the observed delay time distribution (the rate of SNe arising over time from an instantaneous burst of star formation) of Type Ia SNe \citep*[e.g. \citealt{RN1659};][]{RN1317, RN1884}.  If correct, the SD model, together with the lack of agreement with UV fluxes, suggests the secondary stars of progenitor systems are predominately in the lower portion of the proposed {M$_\star \sim 0.7$--$6 $ \MS} range.  

The recent increase in SN discoveries has given astronomers improved data for statistical analysis but has also revealed that over a third of Type Ia SNe are "peculiar" \citep[e.g.][and see \citealt*{RN1664} for a review]{RN1630, RN1325}.  Observations have cast doubt on whether the SD model should continue to be considered the most likely (and only) progenitor model \citep*[e.g. \citealt{RN1495};][]{RN1601}.  Research attempting to constrain progenitors by examining spectra just days after the initial explosion for signs of interaction between the SN ejecta and companion star(s) has generated conflicting results, for example, with the signature of the hydrogen envelope of a main sequence star being detected in some cases, but not in others \citep[e.g.][]{RN1710,RN1709,RN1707}.  Strong limits for the physical size of potential companions have been determined for a small number of cases, with most indicating that a companion, if any, must be compact (e.g. some are limited to 10 to 30 per cent of the Sun's radius, disfavouring RSG companions and also main sequence stars \citep[e.g.][]{RN1732,RN1617,RN1707}.  Searches for surviving companions also continue to be unsuccessful \citep[e.g.][but see \citealt{RN1890}]{RN1863,RN1862}. The study of detailed light curves from the recent Kepler (K2) mission is, however,  providing strong constraints on possible progenitors \citep[e.g.][]{RN1869}. 

Local environmental factors impact on Type Ia SNe,  with recent work revealing that Type Ia SNe located in star forming regions have lower luminosities than their counterparts in locally passive regions \citep{RN1804}.  The colour  of Type Ia SNe  also varies between SNe in the central regions and those in the outskirts of a host galaxy \citep{RN1753}.  The existing global calibrations \citep[e.g.][]{RN1676} can, therefore, lead to bias if these local factors are not taken into account.  
  
Type Ia progenitor models face the challenge of explaining the observed diversity of Type Ia SN events \citep[e.g.][]{RN1826}, while still generating the uniformity and continuous nature of key properties,   \citep[although see][]{RN1734,RN1736}  that makes them such invaluable "standard candles" \citep[][]{RN1680,RN1743,RN1495}.    Progenitor models must explain observations, for example, that brighter/slower Type Ia SNe are only found in star-forming galaxies while elliptical galaxies host fainter cases  \citep[e.g.][]{RN1666,RN1323, RN1650,RN1665}.  The possibility that there are multiple Type Ia SN populations \citep[e.g.][]{RN1672,RN1742, RN1329, RN1324} requiring  two or more progenitor models, is being increasingly considered \citep[e.g.][]{RN1750,RN1634,RN1603,RN1745,RN1748, RN1885}.   The SINGG radial analysis favours long-lived, low mass binaries for at least the majority of Type Ia progenitor systems in star-forming galaxies.

\subsection{Absence of SE SNe in galaxy outskirts}
\label{subsect:centralisation of SE SNe}

The centralisation of SE SNe has previously been linked to the metallicity gradients observed in galaxies \citep[e.g.][]{RN1686,RN1359, RN1691}, with central galactic locations  generally having higher chemical abundances than outer regions  \citep*[e.g.][]{RN1628}.  Metallicity drives the extent of the mass loss from winds for single massive stars,   with low metallicity stars requiring higher masses than their high metallicity counterparts \citep*[e.g.][]{RN1846,RN1546}.     On average, Type Ic SNe are located in regions of higher metallicity  than Type Ib SNe \citep[e.g.][]{RN1511, RN1426} and are often located in the brightest regions of their host galaxies \citep[e.g.][]{RN1641}.  Others find only a weak correlation with metallicity for SE SNe compared to Type II SNe, however, and consider other factors, such as mass, binarity or disturbance, to be of greater importance \citep[e.g.][]{RN1355,RN1426}.   

Compared to other SN types, SE SNe are most associated with leading edges of spiral arms, where compression triggers star formation  \citep[e.g. \citealt* {RN1363};][]{RN1717}.  Galaxy interactions and mergers can also trigger extensive local star formation activity.   SE SN centralisation is more pronounced in disturbed galaxies  \citep[e.g.][]{RN1426,RN1361} and galaxies with very high levels of disturbance have high SE SN to Type II SN ratios \citep{RN1426}.  Galaxy interactions and merger activity reduce metallicity gradients compared to normal isolated galaxies \citep[e.g.][]{RN1756}, implying that metallicity is not the key factor explaining the relative deficiency of SE SNe in the outskirts of galaxies.   

Stellar discs typically have sharp edges when observed using broadband optical wavelengths \citep[\citealt*{RN1870};][and see Fig. \ref{fig:fluxcurves}\textit{a,e,i}]{RN1871,RN1779}  and \HA{} emission \citep[][and see Fig. \ref{fig:fluxcurves}\textit{b,f,j}]{RN433,RN1777}.  Galaxies do not generally exhibit the same sharp truncation in the UV fluxes  as they do in \HA{} \citep[][]{RN485} and many have an XUV disc \citep{RN420}.  The lack of SE SNe in the outskirts of galaxies is consistent with these SNe having higher mass progenitors than Type II SNe and with a reduced massive star formation efficiency (i.e. a bottom-heavy IMF) in the low density, outer regions of galaxies \citep[e.g.][Bruzzese et al. 2019, in prep.]{RN485,RN432,RN1309}.   \citet{RN1665} found  SE SNe under-represented by a factor of about 3 in low mass  galaxies (M < 10$^{10}$ \MS).  Reduced massive SFEs are also found in low \textsc{Hi} mass, low luminosity and low surface brightness galaxies \citep{RN36,RN311,RN322,RN1712, RN1714}.

\section{Conclusions}
\label{sec:Conclusions} 

We have shown that Type Ia SN progenitors in late-type galaxies are best traced by the \textit{R}-band light distribution in the disc, i.e. after allowing for the bulge, and that there is no correlation with UV fluxes.  This is consistent with most, if not all, Type Ia WD progenitors having low mass companions as proposed in the double degenerate model or, in the case of the single degenerate model, with red giant companions \citep{RN1746}.  While the single degenerate model allows for higher mass main sequence sub-giant companions, our results do not support such binary systems  being a major progenitor stream of Type Ia SNe.

The radial distribution of Type II SNe inside R$_\textnormal{{90}}$ (determined using both \textit{R}-band and FUV) is best traced by FUV fluxes.  This is consistent with a growing number of direct detections indicating that Type II SNe have moderately massive progenitors.  

SE SNe have the strongest correlation with the  \HA{} light distribution, supporting the generally held view that  they have the highest mass progenitors (M$_{*}  \gtrsim  ~20$ \MS{}).  It is not consistent, however, with the increasing number of direct detections of moderately massive binary SE SN  progenitors \citep*[but see][]{RN1498}.  At least two distinct SE SN progenitor streams, covering different mass ranges, may be required to explain these conflicting results.

The CCSN population  in the homogeneous SINGG/SUNGG  sample exhibit the well-known centralisation of SE SNe with respect to Type II SNe, with the outskirts of the galaxies being devoid of SE SNe; no SE SN cases occur in the outer third of the FUV fluxes, nor in the outer $\sim20$ per cent of the \textit{R}-band fluxes.  The observations are consistent with high mass progenitors for SE SNe and reduced massive star formation efficiencies in the low density outskirts of galaxies.

\section*{Acknowledgements}

We thank the anonymous referee for helpful and detailed comments that have improved this paper. FAR thanks J.J. Eldridge for useful discussions and comments on an earlier draft of the paper.  Partial funding for the SINGG and SUNGG surveys came from NASA grants NAG5-13083 (LTSA program), GALEX GI04-0105-0009 (NASA GALEX Guest Investigator grant) and NNX09AF85G (GALEX archival grant) to G.R. Meurer.  FAR acknowledges partial funding from the Department of Physics, University of Western Australia, receipt of a Student Travel Award from the Astronomical Society of Australia and financial support from the Space Telescope Science Institute (STScI)  to attend the 2019 STScI Spring Symposium.   This research has made use of the NASA/IPAC Extragalactic Database (NED), which is operated by the Jet Propulsion Laboratory, California Institute of Technology, under contract with the National Aeronautics and Space Administration. The IAU Central Bureau for Astronomical Telegrams database and the Open Supernova Catalog were used to identify SNe that  occurred in the SINGG/SUNGG galaxies.   Anderson-Darling software by SPC for Excel (https://www.spcforexcel.com) was used in this research.

\section*{ORCID iDs}

 Fiona Audcent-Ross		\hspace{4mm} 				https://orcid.org/0000-0002-2770-8004
 
 \noindent{Gerhardt R. Meurer		\hspace{4.5mm}              https://orcid.org/0000-0002-0163-2507}
 
 \noindent{Stuart D. Ryder	 \hspace{9.5mm}		https://orcid.org/0000-0003-4501-8100}
 
\noindent{O. Ivy Wong  		\hspace{13mm}		https://orcid.org/0000-0003-4264-3509}

\noindent{Ji Hoon Kim \hspace{13mm}      	https://orcid.org/0000-0002-1418-3309 }

\bibliographystyle{mnras}    
\bibliography{singgbibsn}   

\begin{thebibliography}{}
\makeatletter
\relax
\def\mn@urlcharsother{\let\do\@makeother \do\$\do\&\do\#\do\^\do\_\do\%\do\~}
\def\mn@doi{\begingroup\mn@urlcharsother \@ifnextchar [ {\mn@doi@}
  {\mn@doi@[]}}
\def\mn@doi@[#1]#2{\def\@tempa{#1}\ifx\@tempa\@empty \href
  {http://dx.doi.org/#2} {doi:#2}\else \href {http://dx.doi.org/#2} {#1}\fi
  \endgroup}
\def\mn@eprint#1#2{\mn@eprint@#1:#2::\@nil}
\def\mn@eprint@arXiv#1{\href {http://arxiv.org/abs/#1} {{\tt arXiv:#1}}}
\def\mn@eprint@dblp#1{\href {http://dblp.uni-trier.de/rec/bibtex/#1.xml}
  {dblp:#1}}
\def\mn@eprint@#1:#2:#3:#4\@nil{\def\@tempa {#1}\def\@tempb {#2}\def\@tempc
  {#3}\ifx \@tempc \@empty \let \@tempc \@tempb \let \@tempb \@tempa \fi \ifx
  \@tempb \@empty \def\@tempb {arXiv}\fi \@ifundefined
  {mn@eprint@\@tempb}{\@tempb:\@tempc}{\expandafter \expandafter \csname
  mn@eprint@\@tempb\endcsname \expandafter{\@tempc}}}

\bibitem[\protect\citeauthoryear{Adams, Kochanek, Gerke, Stanek  \& Dai}{Adams
  et~al.}{2017}]{RN1353}
Adams S.~M.,  Kochanek C.~S.,  Gerke J.~R.,  Stanek K.~Z.,   Dai X.,  2017,
  \mn@doi [MNRAS] {10.1093/mnras/stx816}, 468, 4968

\bibitem[\protect\citeauthoryear{Anderson \& James}{Anderson \&
  James}{2008}]{RN1358}
Anderson J.~P.,  James P.~A.,  2008, \mn@doi [MNRAS]
  {10.1111/j.1365-2966.2008.13843.x}, 390, 1527

\bibitem[\protect\citeauthoryear{Anderson \& James}{Anderson \&
  James}{2009}]{RN1359}
Anderson J.~P.,  James P.~A.,  2009, \mn@doi [MNRAS]
  {10.1111/j.1365-2966.2009.15324.x}, 399, 559

\bibitem[\protect\citeauthoryear{Anderson, Covarrubias, James, Hamuy  \&
  Habergham}{Anderson et~al.}{2010}]{RN1355}
Anderson J.~P.,  Covarrubias R.~A.,  James P.~A.,  Hamuy M.,   Habergham S.~M.,
   2010, \mn@doi [MNRAS] {10.1111/j.1365-2966.2010.17118.x}, 407, 2660

\bibitem[\protect\citeauthoryear{Anderson, Habergham, James  \& Hamuy}{Anderson
  et~al.}{2012}]{RN1357}
Anderson J.~P.,  Habergham S.~M.,  James P.~A.,   Hamuy M.,  2012, \mn@doi
  [MNRAS] {10.1111/j.1365-2966.2012.21324.x}, 424, 1372

\bibitem[\protect\citeauthoryear{Anderson, James, Habergham, Galbany  \&
  Kuncarayakti}{Anderson et~al.}{2015a}]{RN1361}
Anderson J.~P.,  James P.~A.,  Habergham S.~M.,  Galbany L.,   Kuncarayakti H.,
   2015a, \mn@doi [PASA] {10.1017/pasa.2015.19}, 32, 19

\bibitem[\protect\citeauthoryear{Anderson, James, F\"{o}rster,
  Gonz\'{a}lez-Gait\'{a}n, Habergham, Hamuy  \& Lyman}{Anderson
  et~al.}{2015b}]{RN1737}
Anderson J.~P.,  James P.~A.,  F\"{o}rster F.,  Gonz\'{a}lez-Gait\'{a}n S.,
  Habergham S.~M.,  Hamuy M.,   Lyman J.~D.,  2015b, \mn@doi [MNRAS]
  {10.1093/mnras/stu2712}, 448, 732

\bibitem[\protect\citeauthoryear{Anderson et~al.,}{Anderson
  et~al.}{2017}]{RN1724}
Anderson L.~D.,  et~al., 2017, \mn@doi [A\&A] {10.1051/0004-6361/201731019},
  605, 58

\bibitem[\protect\citeauthoryear{Aramyan et~al.,}{Aramyan
  et~al.}{2016}]{RN1363}
Aramyan L.~S.,  et~al., 2016, \mn@doi [MNRAS] {10.1093/mnras/stw873}, 459, 3130

\bibitem[\protect\citeauthoryear{Arcavi et~al.,}{Arcavi et~al.}{2012}]{RN1642}
Arcavi I.,  et~al., 2012, \mn@doi [ApJ] {10.1088/2041-8205/756/2/l30}, 756, L30

\bibitem[\protect\citeauthoryear{Ashall, Mazzali, Sasdelli  \& Prentice}{Ashall
  et~al.}{2016}]{RN1650}
Ashall C.,  Mazzali P.,  Sasdelli M.,   Prentice S.~J.,  2016, \mn@doi [MNRAS]
  {10.1093/mnras/stw1214}, 460, 3529

\bibitem[\protect\citeauthoryear{Audcent-Ross et~al.,}{Audcent-Ross
  et~al.}{2018}]{RN1714}
Audcent-Ross F.~M.,  et~al., 2018, \mn@doi [MNRAS] {10.1093/mnras/sty1538},
  480, 119

\bibitem[\protect\citeauthoryear{Bartunov, Marakova  \& Tsevetkov}{Bartunov
  et~al.}{1992}]{RN1633}
Bartunov O.~S.,  Marakova I.,   Tsevetkov D.,  1992, A\&A, 264, 428

\bibitem[\protect\citeauthoryear{Bartunov, Tsvetkov  \& Pavlyuk}{Bartunov
  et~al.}{2007}]{RN1657}
Bartunov O.~S.,  Tsvetkov D.~Y.,   Pavlyuk N.~N.,  2007, \mn@doi [Highlights
  Astron.] {10.1017/S1743921307010812}, 14, 316

\bibitem[\protect\citeauthoryear{Bastian \& Goodwin}{Bastian \&
  Goodwin}{2006}]{RN1370}
Bastian N.,  Goodwin S.~P.,  2006, \mn@doi [MNRAS]
  {10.1111/j.1745-3933.2006.00162.x}, 369, L9

\bibitem[\protect\citeauthoryear{Bastian, Gieles, Lamers, Scheepmaker  \& de
  Grijs}{Bastian et~al.}{2005}]{RN1823}
Bastian N.,  Gieles M.,  Lamers H.,  Scheepmaker R.,   de Grijs R.,  2005,
  \mn@doi [A\&A] {10.1051/0004-6361:20041078}, 431, 905

\bibitem[\protect\citeauthoryear{Bear \& Soker}{Bear \& Soker}{2018}]{RN1748}
Bear E.,  Soker N.,  2018, \mn@doi [MNRAS] {10.1093/mnras/sty2086}, 480, 3702

\bibitem[\protect\citeauthoryear{Begelman \& Sarazin}{Begelman \&
  Sarazin}{1986}]{RN1764}
Begelman M.~C.,  Sarazin C.~L.,  1986, \mn@doi [ApJ] {10.1086/184637}, 302, L59

\bibitem[\protect\citeauthoryear{Bekki}{Bekki}{2008}]{RN1821}
Bekki K.,  2008, \mn@doi [MNRAS] {10.1111/j.1745-3933.2008.00489.x}, 388, 10

\bibitem[\protect\citeauthoryear{Bellm et~al.,}{Bellm et~al.}{2019}]{RN1868}
Bellm E.~C.,  et~al., 2019, \mn@doi [PASP] {10.1088/1538-3873/aaecbe}, 131,
  018002

\bibitem[\protect\citeauthoryear{Bersten et~al.,}{Bersten
  et~al.}{2018}]{RN1598}
Bersten M.~C.,  et~al., 2018, \mn@doi [Nature] {10.1038/nature25151}, 554, 497

\bibitem[\protect\citeauthoryear{Bethe, Brown, Applegate  \& Lattimer}{Bethe
  et~al.}{1979}]{RN1372}
Bethe H.~A.,  Brown G.~E.,  Applegate J.,   Lattimer J.~M.,  1979, \mn@doi
  [Nucl. Phys. A] {10.1016/0375-9474(79)90596-7}, 324, 487

\bibitem[\protect\citeauthoryear{Blondin et~al.,}{Blondin
  et~al.}{2012}]{RN1826}
Blondin S.,  et~al., 2012, \mn@doi [AJ] {10.1088/0004-6256/143/5/126}, 143, 126

\bibitem[\protect\citeauthoryear{Bloom et~al.,}{Bloom et~al.}{2012a}]{RN1378}
Bloom J.~S.,  et~al., 2012a, \mn@doi [PASP] {10.1086/668468}, 124, 1175

\bibitem[\protect\citeauthoryear{Bloom et~al.,}{Bloom et~al.}{2012b}]{RN1732}
Bloom J.~S.,  et~al., 2012b, \mn@doi [ApJ] {10.1088/2041-8205/744/2/l17}, 744,
  L17

\bibitem[\protect\citeauthoryear{Botticella, Smartt, Kennicutt, Cappellaro,
  Sereno  \& Lee}{Botticella et~al.}{2012}]{RN1382}
Botticella M.~T.,  Smartt S.~J.,  Kennicutt R.~C.,  Cappellaro E.,  Sereno M.,
   Lee J.~C.,  2012, \mn@doi [A\&A] {10.1051/0004-6361/201117343}, 537, A132

\bibitem[\protect\citeauthoryear{Bressert et~al.,}{Bressert
  et~al.}{2010}]{RN1848}
Bressert E.,  et~al., 2010, \mn@doi [MNRAS] {10.1111/j.1745-3933.2010.00946.x},
  409, L54

\bibitem[\protect\citeauthoryear{Brown, Dawson, Harris, Olmstead, Milne  \&
  Roming}{Brown et~al.}{2012}]{RN1709}
Brown P.~J.,  Dawson K.~S.,  Harris D.~W.,  Olmstead M.,  Milne P.,   Roming P.
  W.~A.,  2012, \mn@doi [ApJ] {10.1088/0004-637x/749/1/18}, 749, 18

\bibitem[\protect\citeauthoryear{Bruzzese, Meurer, Lagos, Elson, Werk,
  Blakeslee  \& Ford}{Bruzzese et~al.}{2015}]{RN432}
Bruzzese S.~M.,  Meurer G.~R.,  Lagos C. D.~P.,  Elson E.~C.,  Werk J.~K.,
  Blakeslee J.~P.,   Ford H.,  2015, \mn@doi [MNRAS] {10.1093/mnras/stu2461},
  447, 618

\bibitem[\protect\citeauthoryear{Bruzzese et~al.,}{Bruzzese
  et~al.}{2019}]{RN1894}
Bruzzese S.~M.,  et~al., 2019, MNRAS, in press

\bibitem[\protect\citeauthoryear{Bushouse}{Bushouse}{1987}]{RN1822}
Bushouse H.,  1987, \mn@doi [ApJ] {10.1086/165523}, 320, 49

\bibitem[\protect\citeauthoryear{Canals, Torres  \& Soker}{Canals
  et~al.}{2018}]{RN1601}
Canals P.,  Torres S.,   Soker N.,  2018, \mn@doi [MNRAS]
  {10.1093/mnras/sty2121}, 480, 4519

\bibitem[\protect\citeauthoryear{Cappellaro, Turatto, Tsvetkov, Bartunov,
  Pollas, Evans  \& Hamuy}{Cappellaro et~al.}{1997}]{RN1388}
Cappellaro E.,  Turatto M.,  Tsvetkov D.~Y.,  Bartunov O.~S.,  Pollas C.,
  Evans R.,   Hamuy M.,  1997, A\&A, 322, 431

\bibitem[\protect\citeauthoryear{Cappellaro, Evans  \& Turatto}{Cappellaro
  et~al.}{1999}]{RN1387}
Cappellaro E.,  Evans R.,   Turatto M.,  1999, A\&A, 351, 459

\bibitem[\protect\citeauthoryear{Chakrabarti, Dell, Graur, Filippenko, Lewis
  \& McKee}{Chakrabarti et~al.}{2018}]{RN1892}
Chakrabarti S.,  Dell B.,  Graur O.,  Filippenko A.~V.,  Lewis B.~T.,   McKee
  C.~F.,  2018, \mn@doi [ApJ] {10.3847/2041-8213/aad0a4}, 863

\bibitem[\protect\citeauthoryear{Chandrasekhar}{Chandrasekhar}{1931}]{RN1312}
Chandrasekhar S.,  1931, \mn@doi [ApJ] {10.1086/143324}, 74, 81

\bibitem[\protect\citeauthoryear{Chini, Hoffmeister, Nasseri, Stahl  \&
  Zinnecker}{Chini et~al.}{2012}]{RN1390}
Chini R.,  Hoffmeister V.~H.,  Nasseri A.,  Stahl O.,   Zinnecker H.,  2012,
  \mn@doi [MNRAS] {10.1111/j.1365-2966.2012.21317.x}, 424, 1925

\bibitem[\protect\citeauthoryear{Crockett et~al.,}{Crockett
  et~al.}{2007}]{RN1849}
Crockett R.~M.,  et~al., 2007, \mn@doi [MNRAS]
  {10.1111/j.1365-2966.2007.12283.x}, 381, 835

\bibitem[\protect\citeauthoryear{Crowther}{Crowther}{2007}]{RN1647}
Crowther P.~A.,  2007, \mn@doi [\araa]
  {10.1146/annurev.astro.45.051806.110615}, 45, 177

\bibitem[\protect\citeauthoryear{D{'}Agostino \& Stephens}{D{'}Agostino \&
  Stephens}{1986}]{RN1893}
D{'}Agostino R.,  Stephens M.,  1986, Goodness-of-fit techniques.
Statistics: textbooks and monographs, M. Dekker, New York

\bibitem[\protect\citeauthoryear{Dallaporta}{Dallaporta}{1973}]{RN1672}
Dallaporta N.,  1973, A\&A, 29, 393

\bibitem[\protect\citeauthoryear{{Della Valle} \& Livio}{{Della Valle} \&
  Livio}{1994}]{RN1742}
{Della Valle} M.,  Livio M.,  1994, \mn@doi [ApJ] {10.1086/187228}, 423, L31

\bibitem[\protect\citeauthoryear{Dessart, Hillier, Livne, Yoon, Woosley,
  Waldman  \& Langer}{Dessart et~al.}{2011}]{RN1396}
Dessart L.,  Hillier D.~J.,  Livne E.,  Yoon S.-C.,  Woosley S.,  Waldman R.,
  Langer N.,  2011, \mn@doi [MNRAS] {10.1111/j.1365-2966.2011.18598.x}, 414,
  2985

\bibitem[\protect\citeauthoryear{Dimitriadis et~al.,}{Dimitriadis
  et~al.}{2018}]{RN1869}
Dimitriadis G.,  et~al., 2018, \mn@doi [ApJ] {10.3847/2041-8213/aaedb0}, 870,
  L1

\bibitem[\protect\citeauthoryear{Driver et~al.,}{Driver et~al.}{2006}]{RN1817}
Driver S.,  et~al., 2006, \mn@doi [MNRAS] {10.1111/j.1365-2966.2006.10126.x},
  368, 414

\bibitem[\protect\citeauthoryear{Driver et~al.,}{Driver et~al.}{2018}]{RN687}
Driver S.~P.,  et~al., 2018, \mn@doi [MNRAS] {10.1093/mnras/stx2728}, 475, 2891

\bibitem[\protect\citeauthoryear{Eldridge, Izzard  \& Tout}{Eldridge
  et~al.}{2008}]{RN1648}
Eldridge J.~J.,  Izzard R.~G.,   Tout C.~A.,  2008, \mn@doi [MNRAS]
  {10.1111/j.1365-2966.2007.12738.x}, 384, 1109

\bibitem[\protect\citeauthoryear{Eldridge, Fraser, Smartt, Maund  \&
  Crockett}{Eldridge et~al.}{2013}]{RN1400}
Eldridge J.~J.,  Fraser M.,  Smartt S.~J.,  Maund J.~R.,   Crockett R.~M.,
  2013, \mn@doi [MNRAS] {10.1093/mnras/stt1612}, 436, 774

\bibitem[\protect\citeauthoryear{Eldridge, Fraser, Maund  \& Smartt}{Eldridge
  et~al.}{2015}]{RN1399}
Eldridge J.~J.,  Fraser M.,  Maund J.~R.,   Smartt S.~J.,  2015, \mn@doi
  [MNRAS] {10.1093/mnras/stu2197}, 446, 2689

\bibitem[\protect\citeauthoryear{Filippenko}{Filippenko}{1997}]{RN1404}
Filippenko A.~V.,  1997, \mn@doi [\araa] {10.1146/annurev.astro.35.1.309}, 35,
  309

\bibitem[\protect\citeauthoryear{Filippenko, Matheson  \& Ho}{Filippenko
  et~al.}{1993}]{RN1625}
Filippenko A.~V.,  Matheson T.,   Ho L.~C.,  1993, \mn@doi [ApJ]
  {10.1086/187043}, 415, L103

\bibitem[\protect\citeauthoryear{F\"{o}rster \& Schawinski}{F\"{o}rster \&
  Schawinski}{2008}]{RN1651}
F\"{o}rster F.,  Schawinski K.,  2008, \mn@doi [MNRAS]
  {10.1111/j.1745-3933.2008.00502.x}, 388, L74

\bibitem[\protect\citeauthoryear{Fraser et~al.,}{Fraser et~al.}{2014}]{RN1407}
Fraser M.,  et~al., 2014, \mn@doi [MNRAS] {10.1093/mnrasl/slt179}, 439, L56

\bibitem[\protect\citeauthoryear{Freeman}{Freeman}{1970}]{RN1787}
Freeman K.~C.,  1970, \mn@doi [ApJ] {10.1086/150474}, 160, 811

\bibitem[\protect\citeauthoryear{Fryer}{Fryer}{1999}]{RN1775}
Fryer C.~L.,  1999, \mn@doi [ApJ] {10.1086/307647}, 522, 413

\bibitem[\protect\citeauthoryear{Fryer et~al.,}{Fryer et~al.}{2010}]{RN1747}
Fryer C.~L.,  et~al., 2010, \mn@doi [ApJ] {10.1088/0004-637x/725/1/296}, 725,
  296

\bibitem[\protect\citeauthoryear{Gal-Yam}{Gal-Yam}{2016}]{RN1840}
Gal-Yam A.,  2016, arXiv:, 1611.09353

\bibitem[\protect\citeauthoryear{Gal-Yam, Mazzali, Manulis  \& Bishop}{Gal-Yam
  et~al.}{2013}]{RN1410}
Gal-Yam A.,  Mazzali P.~A.,  Manulis I.,   Bishop D.,  2013, \mn@doi [PASP]
  {10.1086/671483}, 125, 749

\bibitem[\protect\citeauthoryear{Galbany et~al.,}{Galbany
  et~al.}{2014}]{RN1413}
Galbany L.,  et~al., 2014, \mn@doi [A\&A] {10.1051/0004-6361/201424717}, 572,
  38

\bibitem[\protect\citeauthoryear{Gall et~al.,}{Gall et~al.}{2018}]{RN1736}
Gall C.,  et~al., 2018, \mn@doi [A\&A] {10.1051/0004-6361/201730886}, 611, 58

\bibitem[\protect\citeauthoryear{Gaskell, Cappellaro, Dinerstein, Garnett,
  Harkness  \& Wheeler}{Gaskell et~al.}{1986}]{RN1765}
Gaskell C.~M.,  Cappellaro E.,  Dinerstein H.~L.,  Garnett D.~R.,  Harkness
  R.~P.,   Wheeler J.~C.,  1986, \mn@doi [ApJ] {10.1086/184709}, 306, L77

\bibitem[\protect\citeauthoryear{Gil~de Paz et~al.,}{Gil~de Paz
  et~al.}{2005}]{RN484}
Gil~de Paz A.,  et~al., 2005, \mn@doi [ApJ] {10.1086/432054}, 627, L29

\bibitem[\protect\citeauthoryear{Gogarten, Dalcanton, Murphy, Williams, Gilbert
   \& Dolphin}{Gogarten et~al.}{2009}]{RN1417}
Gogarten S.~M.,  Dalcanton J.~J.,  Murphy J.~W.,  Williams B.~F.,  Gilbert K.,
   Dolphin A.,  2009, \mn@doi [ApJ] {10.1088/0004-637x/703/1/300}, 703, 300

\bibitem[\protect\citeauthoryear{Goldstein \& Kasen}{Goldstein \&
  Kasen}{2018}]{RN1662}
Goldstein D.~A.,  Kasen D.,  2018, \mn@doi [ApJ] {10.3847/2041-8213/aaa409},
  852, L33

\bibitem[\protect\citeauthoryear{Gonzalez-Gaitan et~al.,}{Gonzalez-Gaitan
  et~al.}{2015}]{RN1419}
Gonzalez-Gaitan S.,  et~al., 2015, \mn@doi [MNRAS] {10.1093/mnras/stv1097},
  451, 2212

\bibitem[\protect\citeauthoryear{Graur, Bianco, Modjaz, Shivvers, Filippenko,
  Li  \& Smith}{Graur et~al.}{2017}]{RN1665}
Graur O.,  Bianco F.~B.,  Modjaz M.,  Shivvers I.,  Filippenko A.~V.,  Li W.,
  Smith N.,  2017, \mn@doi [ApJ] {10.3847/1538-4357/aa5eb7}, 837, 121

\bibitem[\protect\citeauthoryear{Groh, Meynet  \& Ekstr\"{o}m}{Groh
  et~al.}{2013a}]{RN1420}
Groh J.~H.,  Meynet G.,   Ekstr\"{o}m S.,  2013a, \mn@doi [A\&A]
  {10.1051/0004-6361/201220741}, 550, L7

\bibitem[\protect\citeauthoryear{Groh, Meynet, Georgy  \& Ekstr\"{o}m}{Groh
  et~al.}{2013b}]{RN1421}
Groh J.~H.,  Meynet G.,  Georgy C.,   Ekstr\"{o}m S.,  2013b, \mn@doi [A\&A]
  {10.1051/0004-6361/201321906}, 558, 131

\bibitem[\protect\citeauthoryear{Guillochon, Parrent, Kelley  \&
  Margutti}{Guillochon et~al.}{2017}]{RN1422}
Guillochon J.,  Parrent J.,  Kelley L.~Z.,   Margutti R.,  2017, \mn@doi [ApJ]
  {10.3847/1538-4357/835/1/64}, 835, 64

\bibitem[\protect\citeauthoryear{Habergham, James  \& Anderson}{Habergham
  et~al.}{2012}]{RN1426}
Habergham S.~M.,  James P.~A.,   Anderson J.~P.,  2012, \mn@doi [MNRAS]
  {10.1111/j.1365-2966.2012.21420.x}, 424, 2841

\bibitem[\protect\citeauthoryear{Hachisu, Kato, Saio  \& Nomoto}{Hachisu
  et~al.}{2012}]{RN1739}
Hachisu I.,  Kato M.,  Saio H.,   Nomoto K.,  2012, \mn@doi [ApJ]
  {10.1088/0004-637x/744/1/69}, 744, 69

\bibitem[\protect\citeauthoryear{Hakobyan et~al.,}{Hakobyan
  et~al.}{2016a}]{RN1429}
Hakobyan A.~A.,  et~al., 2016a, \mn@doi [MNRAS] {10.1093/mnras/stv2853}, 456,
  2848

\bibitem[\protect\citeauthoryear{Hakobyan et~al.,}{Hakobyan
  et~al.}{2016b}]{RN1431}
Hakobyan A.~A.,  et~al., 2016b, arXiv:, 1609.08319

\bibitem[\protect\citeauthoryear{Hakobyan et~al.,}{Hakobyan
  et~al.}{2017}]{RN1428}
Hakobyan A.~A.,  et~al., 2017, \mn@doi [MNRAS] {10.1093/mnras/stx1608}, 471,
  1390

\bibitem[\protect\citeauthoryear{Hamuy, Phillips, Maza, Suntzeff, Schommer  \&
  Aviles}{Hamuy et~al.}{1995}]{RN1666}
Hamuy M.,  Phillips M.~M.,  Maza J.,  Suntzeff N.~B.,  Schommer R.~A.,   Aviles
  R.,  1995, \mn@doi [AJ] {10.1086/117251}, 109, 1

\bibitem[\protect\citeauthoryear{Hamuy, Phillips, Suntzeff, Schommer, Maza  \&
  Aviles}{Hamuy et~al.}{1996}]{RN1676}
Hamuy M.,  Phillips M.~M.,  Suntzeff N.~B.,  Schommer R.~A.,  Maza J.,   Aviles
  R.,  1996, \mn@doi [AJ] {10.1086/118190}, 112, 2391

\bibitem[\protect\citeauthoryear{Han \& Podsiadlowski}{Han \&
  Podsiadlowski}{2004}]{RN1704}
Han Z.,  Podsiadlowski P.,  2004, \mn@doi [MNRAS]
  {10.1111/j.1365-2966.2004.07713.x}, 350, 1301

\bibitem[\protect\citeauthoryear{Hanish et~al.,}{Hanish et~al.}{2006}]{RN18}
Hanish D.~J.,  et~al., 2006, \mn@doi [ApJ] {10.1086/504681}, 649, 150

\bibitem[\protect\citeauthoryear{Hayden et~al.,}{Hayden et~al.}{2010}]{RN1710}
Hayden B.~T.,  et~al., 2010, \mn@doi [ApJ] {10.1088/0004-637x/722/2/1691}, 722,
  1691

\bibitem[\protect\citeauthoryear{Heger, Jeannin, Langer  \& Baraffe}{Heger
  et~al.}{1997}]{RN1770}
Heger A.,  Jeannin L.,  Langer N.,   Baraffe I.,  1997, A\&A, 327, 224

\bibitem[\protect\citeauthoryear{Heger, Fryer, Woosley, Langer  \&
  Hartmann}{Heger et~al.}{2003}]{RN1439}
Heger A.,  Fryer C.~L.,  Woosley S.~E.,  Langer N.,   Hartmann D.~H.,  2003,
  \mn@doi [ApJ] {10.1086/375341}, 591, 288

\bibitem[\protect\citeauthoryear{Henry \& Worthey}{Henry \&
  Worthey}{1999}]{RN1628}
Henry R.,  Worthey G.,  1999, \mn@doi [PASP] {10.1086/316403}, 111, 919

\bibitem[\protect\citeauthoryear{Heringer, Pritchet  \& van Kerkwijk}{Heringer
  et~al.}{2019}]{RN1884}
Heringer E.,  Pritchet C.,   van Kerkwijk M.~H.,  2019, \mn@doi [ApJ] {ARTN 52
  10.3847/1538-4357/ab32dd}, 882, 52

\bibitem[\protect\citeauthoryear{Hillebrandt \& Niemeyer}{Hillebrandt \&
  Niemeyer}{2000}]{RN1752}
Hillebrandt W.,  Niemeyer J.~C.,  2000, \mn@doi [\araa]
  {10.1146/annurev.astro.38.1.191}, 38, 191

\bibitem[\protect\citeauthoryear{Horiuchi, Beacom, Kochanek, Prieto, Stanek  \&
  Thompson}{Horiuchi et~al.}{2011}]{RN1447}
Horiuchi S.,  Beacom J.~F.,  Kochanek C.~S.,  Prieto J.~L.,  Stanek K.~Z.,
  Thompson T.~A.,  2011, \mn@doi [ApJ] {10.1088/0004-637x/738/2/154}, 738, 154

\bibitem[\protect\citeauthoryear{Hoversten \& Glazebrook}{Hoversten \&
  Glazebrook}{2008}]{RN311}
Hoversten E.~A.,  Glazebrook K.,  2008, \mn@doi [ApJ] {10.1086/524095}, 675,
  163

\bibitem[\protect\citeauthoryear{Howell}{Howell}{2001}]{RN1329}
Howell D.~A.,  2001, \mn@doi [ApJ] {10.1086/321702}, 554, L193

\bibitem[\protect\citeauthoryear{Howell}{Howell}{2011}]{RN1743}
Howell D.~A.,  2011, \mn@doi [Nature Commun.] {10.1038/ncomms1344}, 2, 350

\bibitem[\protect\citeauthoryear{Howell et~al.,}{Howell et~al.}{2006}]{RN1734}
Howell D.~A.,  et~al., 2006, \mn@doi [Nature] {10.1038/nature05103}, 443, 308

\bibitem[\protect\citeauthoryear{Hoyle \& Fowler}{Hoyle \&
  Fowler}{1960}]{RN1330}
Hoyle F.,  Fowler W.~A.,  1960, \mn@doi [ApJ] {10.1086/146963}, 132, 565

\bibitem[\protect\citeauthoryear{Hygate, Kruijssen, Chevance, Schruba, Haydon
  \& Longmore}{Hygate et~al.}{2019}]{RN1881}
Hygate A. P.~S.,  Kruijssen J. M.~D.,  Chevance M.,  Schruba A.,  Haydon D.~T.,
    Longmore S.~N.,  2019, \mn@doi [MNRAS] {10.1093/mnras/stz1779}, 488, 2800

\bibitem[\protect\citeauthoryear{Iben \& Tutukov}{Iben \&
  Tutukov}{1984}]{RN1314}
Iben I. J.,  Tutukov A.~V.,  1984, \mn@doi [ApJS] {10.1086/190932}, 54, 335

\bibitem[\protect\citeauthoryear{Ivanov, Hamuy  \& Pinto}{Ivanov
  et~al.}{2000}]{RN1655}
Ivanov V.~D.,  Hamuy M.,   Pinto P.~A.,  2000, \mn@doi [ApJ] {10.1086/317060},
  542, 588

\bibitem[\protect\citeauthoryear{James \& Anderson}{James \&
  Anderson}{2006}]{RN1306}
James P.~A.,  Anderson J.~P.,  2006, \mn@doi [A\&A]
  {10.1051/0004-6361:20054509}, 453, 57

\bibitem[\protect\citeauthoryear{Jencson et~al.,}{Jencson
  et~al.}{2019}]{RN1861}
Jencson J.~E.,  et~al., 2019, arXiv:, 1901.00871

\bibitem[\protect\citeauthoryear{Jennings, Williams, Murphy, Dalcanton,
  Gilbert, Dolphin, Weisz  \& Fouesneau}{Jennings et~al.}{2014}]{RN1451}
Jennings Z.~G.,  Williams B.~F.,  Murphy J.~W.,  Dalcanton J.~J.,  Gilbert
  K.~M.,  Dolphin A.~E.,  Weisz D.~R.,   Fouesneau M.,  2014, \mn@doi [ApJ]
  {10.1088/0004-637x/795/2/170}, 795, 170

\bibitem[\protect\citeauthoryear{Jerkstrand, Ergon, Smartt, Fransson,
  Sollerman, Taubenberger, Bersten  \& Spyromilio}{Jerkstrand
  et~al.}{2015}]{RN1754}
Jerkstrand A.,  Ergon M.,  Smartt S.~J.,  Fransson C.,  Sollerman J.,
  Taubenberger S.,  Bersten M.,   Spyromilio J.,  2015, \mn@doi [A\&A]
  {10.1051/0004-6361/201423983}, 573, 12

\bibitem[\protect\citeauthoryear{Johnson \& MacLeod}{Johnson \&
  MacLeod}{1963}]{RN1718}
Johnson H.~M.,  MacLeod J.~M.,  1963, \mn@doi [PASP] {10.1086/127915}, 75, 123

\bibitem[\protect\citeauthoryear{Karapetyan, Hakobyan, Barkhudaryan, Mamon,
  Kunth, Adibekyan  \& Turatto}{Karapetyan et~al.}{2018}]{RN1717}
Karapetyan A.~G.,  Hakobyan A.~A.,  Barkhudaryan L.~V.,  Mamon G.~A.,  Kunth
  D.,  Adibekyan V.,   Turatto M.,  2018, \mn@doi [MNRAS]
  {10.1093/mnras/sty2291}, 481, 566

\bibitem[\protect\citeauthoryear{Kelly, Kirshner  \& Pahre}{Kelly
  et~al.}{2008}]{RN1641}
Kelly P.~L.,  Kirshner R.~P.,   Pahre M.,  2008, \mn@doi [ApJ]
  {10.1086/591925}, 687, 1201

\bibitem[\protect\citeauthoryear{Kenney, Tal, Crowl, Feldmeier  \&
  Jacoby}{Kenney et~al.}{2008}]{RN1814}
Kenney J. D.~P.,  Tal T.,  Crowl H.~H.,  Feldmeier J.,   Jacoby G.~H.,  2008,
  \mn@doi [ApJ] {10.1086/593300}, 687, L69

\bibitem[\protect\citeauthoryear{Kennicutt}{Kennicutt}{1989}]{RN433}
Kennicutt R.~C. J.,  1989, \mn@doi [ApJ] {10.1086/167834}, 344, 685

\bibitem[\protect\citeauthoryear{Kerzendorf, Strampelli, Shen, Schwab, Pakmor,
  Do, Buchner  \& Rest}{Kerzendorf et~al.}{2018}]{RN1862}
Kerzendorf W.~E.,  Strampelli G.,  Shen K.~J.,  Schwab J.,  Pakmor R.,  Do T.,
  Buchner J.,   Rest A.,  2018, \mn@doi [MNRAS] {10.1093/mnras/sty1357}, 479,
  192

\bibitem[\protect\citeauthoryear{Kewley, Rupke, Jabran~Zahid, Geller  \&
  Barton}{Kewley et~al.}{2010}]{RN1756}
Kewley L.~J.,  Rupke D.,  Jabran~Zahid H.,  Geller M.~J.,   Barton E.~J.,
  2010, \mn@doi [ApJ] {10.1088/2041-8205/721/1/l48}, 721, L48

\bibitem[\protect\citeauthoryear{Kilpatrick et~al.,}{Kilpatrick
  et~al.}{2018}]{RN1806}
Kilpatrick C.~D.,  et~al., 2018, \mn@doi [MNRAS] {10.1093/mnras/sty2022}, 480,
  2072

\bibitem[\protect\citeauthoryear{Kochanek}{Kochanek}{2014}]{RN1463}
Kochanek C.~S.,  2014, \mn@doi [ApJ] {10.1088/0004-637x/785/1/28}, 785, 28

\bibitem[\protect\citeauthoryear{Kochanek, Beacom, Kistler, Prieto, Stanek,
  Thompson  \& Y\"{u}ksel}{Kochanek et~al.}{2008}]{RN1467}
Kochanek C.~S.,  Beacom J.~F.,  Kistler M.~D.,  Prieto J.~L.,  Stanek K.~Z.,
  Thompson T.~A.,   Y\"{u}ksel H.,  2008, \mn@doi [ApJ] {10.1086/590053}, 684,
  1336

\bibitem[\protect\citeauthoryear{Kool et~al.,}{Kool et~al.}{2018}]{RN1858}
Kool E.~C.,  et~al., 2018, \mn@doi [MNRAS] {10.1093/mnras/stx2463}, 473, 5641

\bibitem[\protect\citeauthoryear{Koribalski et~al.,}{Koribalski
  et~al.}{2004}]{RN320}
Koribalski B.~S.,  et~al., 2004, \mn@doi [AJ] {10.1086/421744}, 128, 16

\bibitem[\protect\citeauthoryear{Kotak \& Vink}{Kotak \& Vink}{2006}]{RN1470}
Kotak R.,  Vink J.~S.,  2006, \mn@doi [A\&A] {10.1051/0004-6361:20065800}, 460,
  L5

\bibitem[\protect\citeauthoryear{Kregel, Van Der~Kruit  \& Grijs}{Kregel
  et~al.}{2002}]{RN1870}
Kregel M.,  Van Der~Kruit P.~C.,   Grijs R.~d.,  2002, \mn@doi [MNRAS]
  {10.1046/j.1365-8711.2002.05556.x}, 334, 646

\bibitem[\protect\citeauthoryear{Kuncarayakti et~al.,}{Kuncarayakti
  et~al.}{2013a}]{RN1845}
Kuncarayakti H.,  et~al., 2013a, \mn@doi [AJ] {10.1088/0004-6256/146/2/30},
  146, 30

\bibitem[\protect\citeauthoryear{Kuncarayakti et~al.,}{Kuncarayakti
  et~al.}{2013b}]{RN1473}
Kuncarayakti H.,  et~al., 2013b, \mn@doi [AJ] {10.1088/0004-6256/146/2/31},
  146, 31

\bibitem[\protect\citeauthoryear{Kuncarayakti et~al.,}{Kuncarayakti
  et~al.}{2018}]{RN1586}
Kuncarayakti H.,  et~al., 2018, \mn@doi [A\&A] {10.1051/0004-6361/201731923},
  613, 35

\bibitem[\protect\citeauthoryear{Lada \& Lada}{Lada \& Lada}{2003}]{RN451}
Lada C.~J.,  Lada E.~A.,  2003, \mn@doi [\araa]
  {10.1146/annurev.astro.41.011802.094844}, 41, 57

\bibitem[\protect\citeauthoryear{Larson}{Larson}{1976}]{RN1816}
Larson R.~B.,  1976, MNRAS, 176, 31

\bibitem[\protect\citeauthoryear{Leaman, Li, Chornock  \& Filippenko}{Leaman
  et~al.}{2011}]{RN1691}
Leaman J.,  Li W.,  Chornock R.,   Filippenko A.~V.,  2011, \mn@doi [MNRAS]
  {10.1111/j.1365-2966.2011.18158.x}, 412, 1419

\bibitem[\protect\citeauthoryear{Lee \& Lee}{Lee \& Lee}{2014}]{RN1721}
Lee J.~H.,  Lee M.~G.,  2014, \mn@doi [ApJ] {10.1088/0004-637x/786/2/130}, 786,
  130

\bibitem[\protect\citeauthoryear{Lee, Gibson, Flynn, Kawata  \& Beasley}{Lee
  et~al.}{2004}]{RN36}
Lee H.-C.,  Gibson B.~K.,  Flynn C.,  Kawata D.,   Beasley M.~A.,  2004,
  \mn@doi [MNRAS] {10.1111/j.1365-2966.2004.08049.x}, 353, 113

\bibitem[\protect\citeauthoryear{Lee, Kennicutt, {Jos\'{e} G. Funes}, Sakai  \&
  Akiyama}{Lee et~al.}{2009}]{RN1712}
Lee J.~C.,  Kennicutt R.~C.,  {Jos\'{e} G. Funes} S.~J.,  Sakai S.,   Akiyama
  S.,  2009, \mn@doi [ApJ] {10.1088/0004-637x/692/2/1305}, 692, 1305

\bibitem[\protect\citeauthoryear{Lennarz, Altmann  \& Wiebusch}{Lennarz
  et~al.}{2012}]{RN1479}
Lennarz D.,  Altmann D.,   Wiebusch C.,  2012, \mn@doi [A\&A]
  {10.1051/0004-6361/201117666}, 538, A120

\bibitem[\protect\citeauthoryear{Li, Filippenko, Treffers, Riess, Hu  \&
  Qiu}{Li et~al.}{2001}]{RN1630}
Li W.,  Filippenko A.~V.,  Treffers R.~R.,  Riess A.~G.,  Hu J.,   Qiu Y.,
  2001, \mn@doi [ApJ] {10.1086/318299}, 546, 734

\bibitem[\protect\citeauthoryear{Livio \& Mazzali}{Livio \&
  Mazzali}{2018}]{RN1603}
Livio M.,  Mazzali P.,  2018, \mn@doi [Phys. Rep.]
  {10.1016/j.physrep.2018.02.002}, 736, 1

\bibitem[\protect\citeauthoryear{Lovegrove \& Woosley}{Lovegrove \&
  Woosley}{2013}]{RN1487}
Lovegrove E.,  Woosley S.~E.,  2013, \mn@doi [ApJ]
  {10.1088/0004-637x/769/2/109}, 769, 109

\bibitem[\protect\citeauthoryear{Lundqvist et~al.,}{Lundqvist
  et~al.}{2015}]{RN1617}
Lundqvist P.,  et~al., 2015, \mn@doi [A\&A] {10.1051/0004-6361/201525719}, 577,
  39

\bibitem[\protect\citeauthoryear{Lyman, Bersier, James, Mazzali, Eldridge,
  Fraser  \& Pian}{Lyman et~al.}{2016}]{RN1489}
Lyman J.~D.,  Bersier D.,  James P.~A.,  Mazzali P.~A.,  Eldridge J.~J.,
  Fraser M.,   Pian E.,  2016, \mn@doi [MNRAS] {10.1093/mnras/stv2983}, 457,
  328

\bibitem[\protect\citeauthoryear{Lyman et~al.,}{Lyman et~al.}{2018}]{RN1490}
Lyman J.~D.,  et~al., 2018, \mn@doi [MNRAS] {10.1093/mnras/stx2414}, 473, 1359

\bibitem[\protect\citeauthoryear{Magnier et~al.,}{Magnier
  et~al.}{2013}]{RN1725}
Magnier E.~A.,  et~al., 2013, \mn@doi [ApJS] {10.1088/0067-0049/205/2/20}, 205,
  20

\bibitem[\protect\citeauthoryear{Mannucci, Della~Valle, Panagia, Cappellaro,
  Cresci, Maiolino, Petrosian  \& Turatto}{Mannucci et~al.}{2005}]{RN1325}
Mannucci F.,  Della~Valle M.,  Panagia N.,  Cappellaro E.,  Cresci G.,
  Maiolino R.,  Petrosian A.,   Turatto M.,  2005, \mn@doi [A\&A]
  {10.1051/0004-6361:20041411}, 433, 807

\bibitem[\protect\citeauthoryear{Mannucci, Della~Valle  \& Panagia}{Mannucci
  et~al.}{2006}]{RN1324}
Mannucci F.,  Della~Valle M.,   Panagia N.,  2006, \mn@doi [MNRAS]
  {10.1111/j.1365-2966.2006.10501.x}, 370, 773

\bibitem[\protect\citeauthoryear{Maoz, Sharon  \& Gal-Yam}{Maoz
  et~al.}{2010}]{RN1317}
Maoz D.,  Sharon K.,   Gal-Yam A.,  2010, \mn@doi [ApJ]
  {10.1088/0004-637X/722/2/1879}, 722, 1879

\bibitem[\protect\citeauthoryear{Maoz, Mannucci  \& Nelemans}{Maoz
  et~al.}{2014}]{RN1495}
Maoz D.,  Mannucci F.,   Nelemans G.,  2014, \mn@doi [\araa]
  {10.1146/annurev-astro-082812-141031}, 52, 107

\bibitem[\protect\citeauthoryear{Martin \& Kennicutt}{Martin \&
  Kennicutt}{2001}]{RN1777}
Martin C.~L.,  Kennicutt J. R.~C.,  2001, \mn@doi [ApJ] {10.1086/321452}, 555,
  301

\bibitem[\protect\citeauthoryear{Masci et~al.,}{Masci et~al.}{2017}]{RN1726}
Masci F.~J.,  et~al., 2017, \mn@doi [PASP] {10.1088/1538-3873/129/971/014002},
  129, 014002

\bibitem[\protect\citeauthoryear{Mattila et~al.,}{Mattila
  et~al.}{2012}]{RN1497}
Mattila S.,  et~al., 2012, \mn@doi [ApJ] {10.1088/0004-637x/756/2/111}, 756,
  111

\bibitem[\protect\citeauthoryear{Maund}{Maund}{2018}]{RN1498}
Maund J.~R.,  2018, \mn@doi [MNRAS] {10.1093/mnras/sty093}, 476, 2629

\bibitem[\protect\citeauthoryear{Maund, Smartt, Kudritzki, Podsiadlowski  \&
  Gilmore}{Maund et~al.}{2004}]{RN1501}
Maund J.~R.,  Smartt S.~J.,  Kudritzki R.~P.,  Podsiadlowski P.,   Gilmore
  G.~F.,  2004, \mn@doi [Nature] {10.1038/nature02161}, 427, 129

\bibitem[\protect\citeauthoryear{Mazzali, Sauer, Pian, Deng, Prentice, Ben~Ami,
  Taubenberger  \& Nomoto}{Mazzali et~al.}{2017}]{RN1605}
Mazzali P.~A.,  Sauer D.~N.,  Pian E.,  Deng J.,  Prentice S.,  Ben~Ami S.,
  Taubenberger S.,   Nomoto K.,  2017, \mn@doi [MNRAS] {10.1093/mnras/stx992},
  469, 2498

\bibitem[\protect\citeauthoryear{Meurer et~al.,}{Meurer et~al.}{2006}]{RN41}
Meurer G.~R.,  et~al., 2006, \mn@doi [ApJS] {10.1086/504685}, 165, 307

\bibitem[\protect\citeauthoryear{Meurer et~al.,}{Meurer et~al.}{2009}]{RN322}
Meurer G.~R.,  et~al., 2009, \mn@doi [ApJ] {10.1088/0004-637x/695/1/765}, 695,
  765

\bibitem[\protect\citeauthoryear{Meurer, Obreschkow, Wong, Zheng, Audcent-Ross
  \& Hanish}{Meurer et~al.}{2018}]{RN1779}
Meurer G.~R.,  Obreschkow D.,  Wong O.~I.,  Zheng Z.,  Audcent-Ross F.~M.,
  Hanish D.~J.,  2018, \mn@doi [MNRAS] {10.1093/mnras/sty275}, 476, 1624

\bibitem[\protect\citeauthoryear{Meyer et~al.,}{Meyer et~al.}{2004}]{RN321}
Meyer M.~J.,  et~al., 2004, \mn@doi [MNRAS] {10.1111/j.1365-2966.2004.07710.x},
  350, 1195

\bibitem[\protect\citeauthoryear{Minkowski}{Minkowski}{1941}]{RN1813}
Minkowski R.,  1941, \mn@doi [PASP] {10.1086/125315}, 53, 224

\bibitem[\protect\citeauthoryear{Modjaz}{Modjaz}{2012}]{RN1511}
Modjaz M.,  2012, \mn@doi [Proc. IAU] {10.1017/s1743921312012938}, 7, 207

\bibitem[\protect\citeauthoryear{Morselli, Popesso, Erfanianfar  \&
  Concas}{Morselli et~al.}{2017}]{RN1807}
Morselli L.,  Popesso P.,  Erfanianfar G.,   Concas A.,  2017, \mn@doi [A\&A]
  {10.1051/0004-6361/201629409}, 597, 97

\bibitem[\protect\citeauthoryear{Nomoto}{Nomoto}{1982}]{RN1744}
Nomoto K.,  1982, \mn@doi [ApJ] {10.1086/159682}, 253, 798

\bibitem[\protect\citeauthoryear{Nomoto \& Leung}{Nomoto \&
  Leung}{2018}]{RN1746}
Nomoto K.,  Leung S.-C.,  2018, \mn@doi [Space Sci. Rev.]
  {10.1007/s11214-018-0499-0}, 214, 67

\bibitem[\protect\citeauthoryear{Nomoto, Thielemann  \& Yokoi}{Nomoto
  et~al.}{1984}]{RN1705}
Nomoto K.,  Thielemann F.~K.,   Yokoi K.,  1984, \mn@doi [ApJ]
  {10.1086/162639}, 286, 644

\bibitem[\protect\citeauthoryear{Nomoto, Iwamoto  \& Suzuki}{Nomoto
  et~al.}{1995}]{RN1606}
Nomoto K.,  Iwamoto K.,   Suzuki T.,  1995, \mn@doi [Phys. Rep.]
  {10.1016/0370-1573(94)00107-e}, 256, 173

\bibitem[\protect\citeauthoryear{Oemler \& Tinsley}{Oemler \&
  Tinsley}{1979}]{RN1326}
Oemler A. J.,  Tinsley B.~M.,  1979, \mn@doi [AJ] {10.1086/112502}, 84, 985

\bibitem[\protect\citeauthoryear{Patterson}{Patterson}{1940}]{RN1793}
Patterson F.,  1940, Harv. Coll. Obs. Bull., 914, 9

\bibitem[\protect\citeauthoryear{Peletier \& Balcells}{Peletier \&
  Balcells}{1996}]{RN1820}
Peletier R.~F.,  Balcells M.,  1996, \mn@doi [AJ] {10.1086/117958}, 111, 223

\bibitem[\protect\citeauthoryear{Perna, Duffell, Cantiello  \& MacFadyen}{Perna
  et~al.}{2014}]{RN1812}
Perna R.,  Duffell P.,  Cantiello M.,   MacFadyen A.~I.,  2014, \mn@doi [ApJ]
  {10.1088/0004-637x/781/2/119}, 781, 119

\bibitem[\protect\citeauthoryear{Petrosian et~al.,}{Petrosian
  et~al.}{2005}]{RN1686}
Petrosian A.,  et~al., 2005, \mn@doi [AJ] {10.1086/427712}, 129, 1369

\bibitem[\protect\citeauthoryear{Phillips, Lira, Suntzeff, Schommer, Hamuy  \&
  Maza}{Phillips et~al.}{1999}]{RN1679}
Phillips M.~M.,  Lira P.,  Suntzeff N.~B.,  Schommer R.~A.,  Hamuy M.,   Maza
  J.,  1999, \mn@doi [AJ] {10.1086/301032}, 118, 1766

\bibitem[\protect\citeauthoryear{Podsiadlowski, Joss  \& Hsu}{Podsiadlowski
  et~al.}{1992}]{RN1305}
Podsiadlowski P.,  Joss P.~C.,   Hsu J. J.~L.,  1992, \mn@doi [ApJ]
  {10.1086/171341}, 391, 246

\bibitem[\protect\citeauthoryear{Portegies~Zwart, McMillan  \&
  Gieles}{Portegies~Zwart et~al.}{2010}]{RN1847}
Portegies~Zwart S.~F.,  McMillan S. L.~W.,   Gieles M.,  2010, \mn@doi [\araa]
  {10.1146/annurev-astro-081309-130834}, 48, 431

\bibitem[\protect\citeauthoryear{Porter \& Filippenko}{Porter \&
  Filippenko}{1987}]{RN1620}
Porter A.~C.,  Filippenko A.~V.,  1987, \mn@doi [AJ] {10.1086/114420}, 93, 1372

\bibitem[\protect\citeauthoryear{Prentice et~al.,}{Prentice
  et~al.}{2018}]{RN1595}
Prentice S.~J.,  et~al., 2018, \mn@doi [MNRAS] {10.1093/mnras/sty1223}, 478,
  4162

\bibitem[\protect\citeauthoryear{Reynolds, Fraser  \& Gilmore}{Reynolds
  et~al.}{2015}]{RN1815}
Reynolds T.~M.,  Fraser M.,   Gilmore G.,  2015, \mn@doi [MNRAS]
  {10.1093/mnras/stv1809}, 453, 2885

\bibitem[\protect\citeauthoryear{Riess et~al.,}{Riess et~al.}{1998}]{RN1735}
Riess A.~G.,  et~al., 1998, \mn@doi [AJ] {10.1086/300499}, 116, 1009

\bibitem[\protect\citeauthoryear{Rigault et~al.,}{Rigault
  et~al.}{2013}]{RN1804}
Rigault M.,  et~al., 2013, \mn@doi [A\&A] {10.1051/0004-6361/201322104}, 560,
  A66

\bibitem[\protect\citeauthoryear{Roman et~al.,}{Roman et~al.}{2018}]{RN1753}
Roman M.,  et~al., 2018, \mn@doi [A\&A] {10.1051/0004-6361/201731425}, 615, 68

\bibitem[\protect\citeauthoryear{Ro\u{s}kar, Debattista, Stinson, Quinn,
  Kaufmann  \& Wadsley}{Ro\u{s}kar et~al.}{2008}]{RN1780}
Ro\u{s}kar R.,  Debattista V.~P.,  Stinson G.~S.,  Quinn T.~R.,  Kaufmann T.,
  Wadsley J.,  2008, \mn@doi [ApJ] {10.1086/586734}, 675, L65

\bibitem[\protect\citeauthoryear{Ryder et~al.,}{Ryder et~al.}{2018}]{RN1531}
Ryder S.~D.,  et~al., 2018, \mn@doi [ApJ] {10.3847/1538-4357/aaaf1e}, 856, 83

\bibitem[\protect\citeauthoryear{Sana et~al.,}{Sana et~al.}{2012}]{RN1533}
Sana H.,  et~al., 2012, \mn@doi [Science] {10.1126/science.1223344}, 337, 444

\bibitem[\protect\citeauthoryear{Sancisi, Fraternali, Oosterloo  \& van~der
  Hulst}{Sancisi et~al.}{2008}]{RN1808}
Sancisi R.,  Fraternali F.,  Oosterloo T.,   van~der Hulst T.,  2008, \mn@doi
  [{A\&AR}] {10.1007/s00159-008-0010-0}, 15, 189

\bibitem[\protect\citeauthoryear{Sanders et~al.,}{Sanders
  et~al.}{2015}]{RN1535}
Sanders N.~E.,  et~al., 2015, \mn@doi [ApJ] {10.1088/0004-637x/799/2/208}, 799,
  208

\bibitem[\protect\citeauthoryear{Schaefer \& Pagnotta}{Schaefer \&
  Pagnotta}{2012}]{RN1863}
Schaefer B.~E.,  Pagnotta A.,  2012, \mn@doi [Nature] {10.1038/nature10692},
  481, 164

\bibitem[\protect\citeauthoryear{Shappee, Piro, Stanek, Patel, Margutti,
  Lipunov  \& Pogge}{Shappee et~al.}{2018}]{RN1707}
Shappee B.~J.,  Piro A.~L.,  Stanek K.~Z.,  Patel S.~G.,  Margutti R.~A.,
  Lipunov V.~M.,   Pogge R.~W.,  2018, \mn@doi [ApJ]
  {10.3847/1538-4357/aaa1e9}, 855, 6

\bibitem[\protect\citeauthoryear{Shen, Kasen, Miles  \& Townsley}{Shen
  et~al.}{2018a}]{RN1661}
Shen K.~J.,  Kasen D.,  Miles B.~J.,   Townsley D.~M.,  2018a, \mn@doi [ApJ]
  {10.3847/1538-4357/aaa8de}, 854, 52

\bibitem[\protect\citeauthoryear{Shen et~al.,}{Shen et~al.}{2018b}]{RN1890}
Shen K.~J.,  et~al., 2018b, \mn@doi [ApJ] {10.3847/1538-4357/aad55b}, 865, 15

\bibitem[\protect\citeauthoryear{Shivvers et~al.,}{Shivvers
  et~al.}{2017}]{RN1755}
Shivvers I.,  et~al., 2017, \mn@doi [PASP] {10.1088/1538-3873/aa54a6}, 129,
  4201

\bibitem[\protect\citeauthoryear{Smartt}{Smartt}{2015}]{RN1543}
Smartt S.~J.,  2015, \mn@doi [PASA] {10.1017/pasa.2015.17}, 32, 22

\bibitem[\protect\citeauthoryear{Smartt, Gilmore, Tout  \& Hodgkin}{Smartt
  et~al.}{2002}]{RN1545}
Smartt S.~J.,  Gilmore G.~F.,  Tout C.~A.,   Hodgkin S.~T.,  2002, \mn@doi
  [ApJ] {10.1086/324690}, 565, 1089

\bibitem[\protect\citeauthoryear{Smartt, Eldridge, Crockett  \& Maund}{Smartt
  et~al.}{2009}]{RN1544}
Smartt S.~J.,  Eldridge J.~J.,  Crockett R.~M.,   Maund J.~R.,  2009, \mn@doi
  [MNRAS] {10.1111/j.1365-2966.2009.14506.x}, 395, 1409

\bibitem[\protect\citeauthoryear{Smith}{Smith}{1968}]{RN1819}
Smith L.~F.,  1968, \mn@doi [MNRAS] {10.1093/mnras}, 141, 317

\bibitem[\protect\citeauthoryear{Smith}{Smith}{2014}]{RN1546}
Smith N.,  2014, \mn@doi [\araa] {10.1146/annurev-astro-081913-040025}, 52, 487

\bibitem[\protect\citeauthoryear{Smith \& Owocki}{Smith \&
  Owocki}{2006}]{RN1772}
Smith N.,  Owocki S.~P.,  2006, \mn@doi [ApJ] {10.1086/506523}, 645, L45

\bibitem[\protect\citeauthoryear{Smith, Li, Filippenko  \& Chornock}{Smith
  et~al.}{2011a}]{RN1547}
Smith N.,  Li W.,  Filippenko A.~V.,   Chornock R.,  2011a, \mn@doi [MNRAS]
  {10.1111/j.1365-2966.2011.17229.x}, 412, 1522

\bibitem[\protect\citeauthoryear{Smith, Gehrz, Campbell, Kassis, Le~Mignant,
  Kuluhiwa  \& Filippenko}{Smith et~al.}{2011b}]{RN1771}
Smith N.,  Gehrz R.~D.,  Campbell R.,  Kassis M.,  Le~Mignant D.,  Kuluhiwa K.,
    Filippenko A.~V.,  2011b, \mn@doi [MNRAS]
  {10.1111/j.1365-2966.2011.19614.x}, 418, 1959

\bibitem[\protect\citeauthoryear{Smith, Mauerhan  \& Prieto}{Smith
  et~al.}{2014}]{RN1762}
Smith N.,  Mauerhan J.~C.,   Prieto J.~L.,  2014, \mn@doi [MNRAS]
  {10.1093/mnras/stt2269}, 438, 1191

\bibitem[\protect\citeauthoryear{Soker}{Soker}{2019}]{RN1885}
Soker N.,  2019, \mn@doi [MNRAS] {10.1093/mnras/stz2817}, 490, 2430

\bibitem[\protect\citeauthoryear{Stoner}{Stoner}{2011}]{RN1319}
Stoner E.~C.,  2011, \mn@doi [Philos. Mag.] {10.1080/14786435.2011.586407}, 91,
  3423

\bibitem[\protect\citeauthoryear{Stritzinger et~al.,}{Stritzinger
  et~al.}{2018}]{RN1745}
Stritzinger M.~D.,  et~al., 2018, {ApJ}, 864, L35

\bibitem[\protect\citeauthoryear{Sukhbold, Ertl, Woosley, Brown  \&
  Janka}{Sukhbold et~al.}{2016}]{RN1827}
Sukhbold T.,  Ertl T.,  Woosley S.~E.,  Brown J.~M.,   Janka H.~T.,  2016,
  \mn@doi [ApJ] {10.3847/0004-637X/821/1/38}, 821, 38

\bibitem[\protect\citeauthoryear{Sullivan, Treyer, Ellis, Bridges, Milliard  \&
  Donas}{Sullivan et~al.}{2000}]{RN648}
Sullivan M.,  Treyer M.~A.,  Ellis R.~S.,  Bridges T.~J.,  Milliard B.,   Donas
  J.,  2000, \mn@doi [MNRAS] {10.1046/j.1365-8711.2000.03140.x}, 312, 442

\bibitem[\protect\citeauthoryear{Sullivan et~al.,}{Sullivan
  et~al.}{2006}]{RN1323}
Sullivan M.,  et~al., 2006, \mn@doi [ApJ] {10.1086/506137}, 648, 868

\bibitem[\protect\citeauthoryear{Sullivan et~al.,}{Sullivan
  et~al.}{2010}]{RN1680}
Sullivan M.,  et~al., 2010, \mn@doi [MNRAS] {10.1111/j.1365-2966.2010.16731.x},
  406, 782

\bibitem[\protect\citeauthoryear{Taddia et~al.,}{Taddia et~al.}{2018}]{RN1612}
Taddia F.,  et~al., 2018, \mn@doi [A\&A] {10.1051/0004-6361/201730844}, 609,
  136

\bibitem[\protect\citeauthoryear{Taubenberger}{Taubenberger}{2017}]{RN1664}
Taubenberger S.,  2017, The Extremes of Thermonuclear Supernovae.
Springer International Publishing AG, \mn@doi{10.1007/978-3-319-21846-5-37}

\bibitem[\protect\citeauthoryear{Thilker et~al.,}{Thilker et~al.}{2005}]{RN485}
Thilker D.~A.,  et~al., 2005, \mn@doi [ApJ] {10.1086/425251}, 619, L79

\bibitem[\protect\citeauthoryear{Thilker et~al.,}{Thilker et~al.}{2007}]{RN420}
Thilker D.~A.,  et~al., 2007, \mn@doi [ApJS] {10.1086/523853}, 173, 538

\bibitem[\protect\citeauthoryear{Totani, Morokuma, Oda, Doi  \& Yasuda}{Totani
  et~al.}{2008}]{RN1659}
Totani T.,  Morokuma T.,  Oda T.,  Doi M.,   Yasuda N.,  2008, \mn@doi [PASJ]
  {10.1093/pasj/60.6.1327}, 60, 1327

\bibitem[\protect\citeauthoryear{Tsvetkov, Pavlyuk  \& Bartunov}{Tsvetkov
  et~al.}{2004}]{RN1629}
Tsvetkov D.~Y.,  Pavlyuk N.~N.,   Bartunov O.~S.,  2004, \mn@doi [Astron.
  Lett.] {10.1134/1.1819491}, 30, 729

\bibitem[\protect\citeauthoryear{Valenti et~al.,}{Valenti
  et~al.}{2016}]{RN1561}
Valenti S.,  et~al., 2016, \mn@doi [MNRAS] {10.1093/mnras/stw870}, 459, 3939

\bibitem[\protect\citeauthoryear{Van~Dyk}{Van~Dyk}{2017}]{RN1563}
Van~Dyk S.~D.,  2017, \mn@doi [Philos. Trans. Royal Soc.]
  {10.1098/rsta.2016.0277}, 375, 2016277

\bibitem[\protect\citeauthoryear{Van~Dyk et~al.,}{Van~Dyk
  et~al.}{2018}]{RN1597}
Van~Dyk S.~D.,  et~al., 2018, \mn@doi [ApJ] {10.3847/1538-4357/aac32c}, 860, 90

\bibitem[\protect\citeauthoryear{Vink, de Koter  \& Lamers}{Vink
  et~al.}{2001}]{RN1846}
Vink J.~S.,  de Koter A.,   Lamers H. J. G. L.~M.,  2001, \mn@doi [A\&A]
  {10.1051/0004-6361:20010127}, 369, 574

\bibitem[\protect\citeauthoryear{Vu\u{c}eti\'{c}, Arbutina  \&
  Uro\u{s}evi\'{c}}{Vu\u{c}eti\'{c} et~al.}{2015}]{RN1719}
Vu\u{c}eti\'{c} M.~M.,  Arbutina B.,   Uro\u{s}evi\'{c} D.,  2015, \mn@doi
  [MNRAS] {10.1093/mnras/stu2093}, 446, 943

\bibitem[\protect\citeauthoryear{Wang \& Han}{Wang \& Han}{2012}]{RN1750}
Wang B.,  Han Z.,  2012, \mn@doi [New Astron. Rev.]
  {10.1016/j.newar.2012.04.001}, 56, 122

\bibitem[\protect\citeauthoryear{Wang, H\"{o}flich  \& Wheeler}{Wang
  et~al.}{1997}]{RN1627}
Wang L.,  H\"{o}flich P.,   Wheeler J.~C.,  1997, \mn@doi [ApJ]
  {10.1086/310737}, 483, L29

\bibitem[\protect\citeauthoryear{Wang, Wang, Filippenko, Zhang  \& Zhao}{Wang
  et~al.}{2013}]{RN1634}
Wang X.,  Wang L.,  Filippenko A.~V.,  Zhang T.,   Zhao X.,  2013, \mn@doi
  [Science] {10.1126/science.1231502}, 340, 170

\bibitem[\protect\citeauthoryear{Wang et~al.,}{Wang et~al.}{2018}]{RN1782}
Wang J.,  et~al., 2018, \mn@doi [MNRAS] {10.1093/mnras/sty1687}, 479, 4292

\bibitem[\protect\citeauthoryear{Watts, Meurer, Lagos, Bruzzese, Kroupa  \&
  Jerabkova}{Watts et~al.}{2018}]{RN1309}
Watts A.~B.,  Meurer G.~R.,  Lagos C. D.~P.,  Bruzzese S.~M.,  Kroupa P.,
  Jerabkova T.,  2018, \mn@doi [MNRAS] {10.1093/mnras/sty1006}, 477, 5554

\bibitem[\protect\citeauthoryear{Webbink}{Webbink}{1984}]{RN1315}
Webbink R.~F.,  1984, \mn@doi [ApJ] {10.1086/161701}, 277, 355

\bibitem[\protect\citeauthoryear{Whelan \& Iben}{Whelan \& Iben}{1973}]{RN1310}
Whelan J.,  Iben Icko J.,  1973, \mn@doi [ApJ] {10.1086/152565}, 186, 1007

\bibitem[\protect\citeauthoryear{Whipple}{Whipple}{1939}]{RN1715}
Whipple F.~L.,  1939, \mn@doi [Proc. Natl. Acad. Sci. USA]
  {10.1073/pnas.25.3.118}, 25, 118

\bibitem[\protect\citeauthoryear{Williams, Peterson, Murphy, Gilbert,
  Dalcanton, Dolphin  \& Jennings}{Williams et~al.}{2014}]{RN1568}
Williams B.~F.,  Peterson S.,  Murphy J.,  Gilbert K.,  Dalcanton J.~J.,
  Dolphin A.~E.,   Jennings Z.~G.,  2014, \mn@doi [ApJ]
  {10.1088/0004-637x/791/2/105}, 791, 105

\bibitem[\protect\citeauthoryear{Williams, Hillis, Murphy, Gilbert, Dalcanton
  \& Dolphin}{Williams et~al.}{2018}]{RN1593}
Williams B.~F.,  Hillis T.~J.,  Murphy J.~W.,  Gilbert K.,  Dalcanton J.~J.,
  Dolphin A.~E.,  2018, \mn@doi [ApJ] {10.3847/1538-4357/aaba7d}, 860, 39

\bibitem[\protect\citeauthoryear{Wong}{Wong}{2007}]{RN381}
Wong O.~I.,  2007, Phd, Univ. Melbourne

\bibitem[\protect\citeauthoryear{Wong, Meurer, Zheng, Heckman, Thilker  \&
  Zwaan}{Wong et~al.}{2016}]{RN257}
Wong O.~I.,  Meurer G.~R.,  Zheng Z.,  Heckman T.~M.,  Thilker D.~A.,   Zwaan
  M.~A.,  2016, \mn@doi [MNRAS] {10.1093/mnras/stw993}, 460, 1106

\bibitem[\protect\citeauthoryear{Zheng et~al.,}{Zheng et~al.}{2015}]{RN1786}
Zheng Z.,  et~al., 2015, \mn@doi [ApJ] {10.1088/0004-637x/800/2/120}, 800, 120

\bibitem[\protect\citeauthoryear{Zwaan et~al.,}{Zwaan et~al.}{2004}]{RN319}
Zwaan M.~A.,  et~al., 2004, \mn@doi [MNRAS] {10.1111/j.1365-2966.2004.07782.x},
  350, 1210

\bibitem[\protect\citeauthoryear{van~den Bergh}{van~den Bergh}{1997}]{RN1562}
van~den Bergh S.,  1997, \mn@doi [AJ] {10.1086/118244}, 113, 197

\bibitem[\protect\citeauthoryear{{van~der~Kruit}}{{van~der~Kruit}}{2007}]{RN1871}
{van~der~Kruit} P.~C.,  2007, \mn@doi [A\&A] {10.1051/0004-6361:20066941}, 466,
  883

\makeatother
\end{thebibliography}

\appendix

\section{Enclosed flux fraction histograms}
\label{appendix_histograms}

\begin{figure*}
\centering
\hspace*{-1cm}
\includegraphics[scale=0.79]{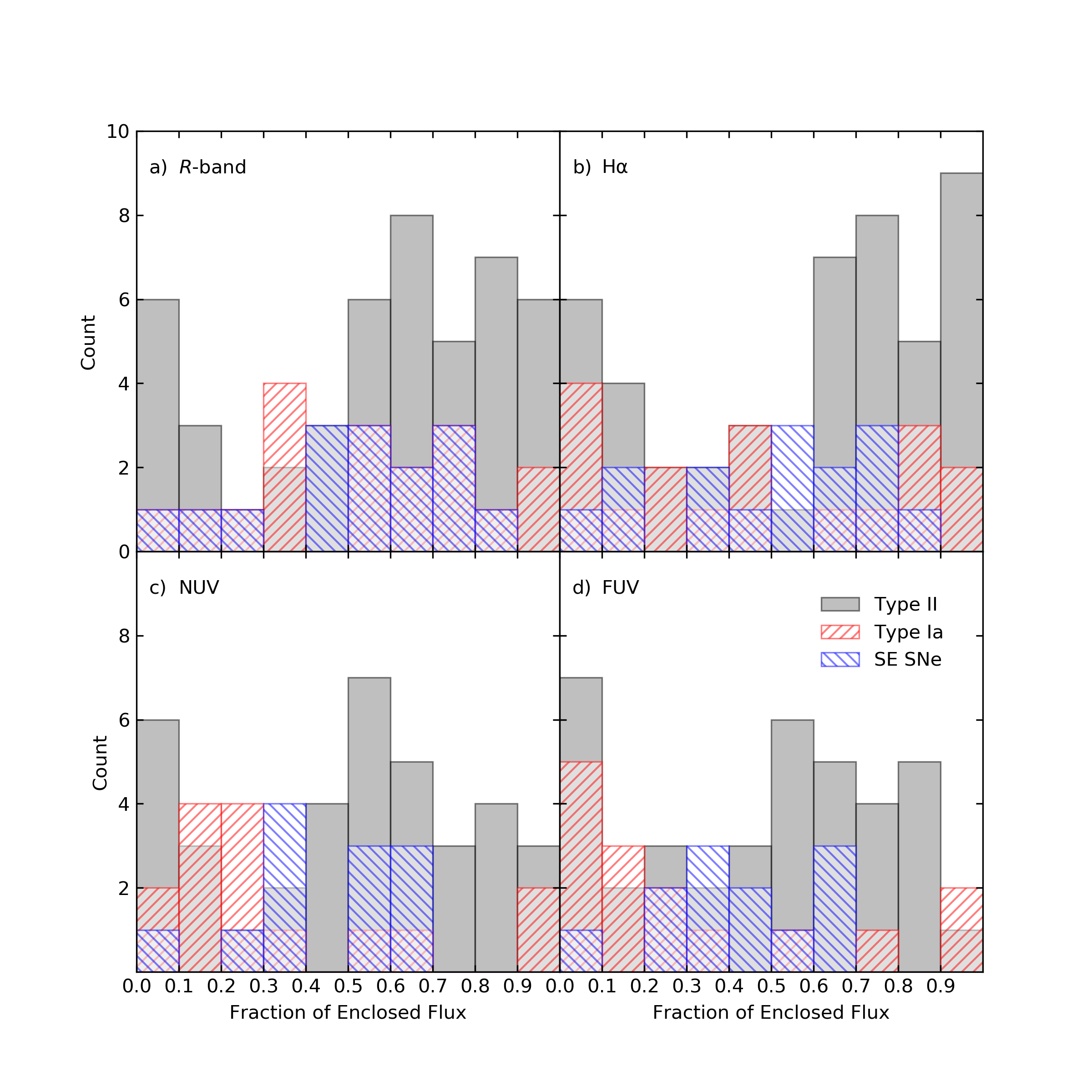}
\caption{Histograms of the radial enclosed flux fractions (see Section \ref{sec:method} for method).   This figure is an alternative representation of the results in Figures \ref{fig:ia_all_duo}\textit{a} and \ref{fig:trioplot}\textit{a,b}, and similarly illustrates, for example, a central deficit of Type Ia SNe and the lack of Type II and SE SNe in the outskirts of the UV flux distributions of our sample.  Type II SNe are indicated with solid grey, Type Ia with red upward-slanting lines, and SE SNe with blue downward-slanting lines (as shown in the key above).}  
\label{fig:fig9999}
\end{figure*}

Figure \ref{fig:fig9999} provides an alternative to the CDF presentation of the radial distribution of SNe used in the body of this paper and reinforces its results.  The lack of Type II and SE SNe in the outskirts of the UV radial flux distributions of our sample is particularly evident in Fig. \ref{fig:fig9999}\textit{c,d}, for example.

The statistical significance of our results using Kolmogorov-Smirnov (KS) testing are detailed in the body of this paper.  The Anderson-Darling (AD) test \citep[case 3,][]{RN1893} was also used to examine the significance of our results  and generated results consistent with the KS testing. The radial distribution of Type II SNe over the full apertures is not consistent with either \textit{R}-band and \HA{} radial fluxes  (Anderson-Darling AD* > $\sim$ 2 and $\textnormal{p}_{AD} = 0.0$), while NUV has the only non-negligible $\textnormal{p}_{AD}$ value (AD* $= 0.6$ and $\textnormal{p}_{AD} = 0.12$).  In statistically significant results, albeit with small numbers of SNe, the radial distribution of the 18 Type Ia SNe over the full apertures is not consistent with UV radial fluxes  (Anderson-Darling {AD* $>$ 3} and $\textnormal{p}_{AD} =0.0$) and is consistent with \textit{R}-band fluxes ({AD*$ = 0.13$} and p$_{AD} = 0.98$), while only \HA{} fluxes may trace the 15 SE SNe (AD* $ = 0.53$ and $\textnormal{p}_{AD} = 0.18$), with all other radial fluxes being statistically rejected (AD* > 1 and $\textnormal{p}_{AD} \leq 0.01$).

\end{document}